\documentclass[preprintnumbers,amsmath,amssymb]{revtex4}
\usepackage{amsmath,amssymb,color,epsfig,latexsym,natbib,graphicx,dcolumn, bm} 
\usepackage{xifthen}
\usepackage{lpic}
\usepackage{caption}
\usepackage{floatrow}
\usepackage{subfigure}
\usepackage[normalem]{ulem}

\bibpunct{[}{]}{,}{n}{,}{,} 

\def\DEL#1{{\textcolor{green}{}}}   
\newcommand{\vomega}{\mbox{\boldmath $\omega$}}    \newcommand{\BV}{{Brunt-V\"ais\"al\"a frequency}}        
\newcommand{\be}{\begin{equation}}  \newcommand{\ee}{\end{equation}}

\begin{document}
\title{Dual constant-flux energy cascades to both large scales and small scales} 

\author{A. Pouquet$^{1,2}$, R. Marino$^{3}$, P.D. Mininni$^4$ and D. Rosenberg$^5$} 

\affiliation{$^1$NCAR, P.O. Box 3000, Boulder, Colorado 80307-3000, USA. \\
$^2$Laboratory for Atmospheric and Space Physics, CU, Boulder, CO, 80309-256 USA.\\
$^3$Laboratoire de M\'{e}canique des Fluides et d'Acoustique, CNRS, \'{E}cole Centrale de Lyon,
Universit\'{e} de Lyon, 69134 \'{E}cully, FRANCE.\\
$^4$Departamento de Fisica, Facultad de Cs. Exactas y Naturales, Universidad de Buenos Aires \\
Ciudad Universitaria, Pabellon 1, Buenos Aires 1428, ARGENTINA. \\
$^5$ Cooperative Institute for Research in the Atmosphere, NOAA, Boulder, CO, 80305 USA.}

\begin{abstract} 
In this paper, we present an overview of concepts and data concerning inverse cascades of excitation towards scales larger than the forcing scale in a variety of contexts, from two-dimensional fluids  and wave turbulence, to geophysical flows in the presence of rotation and stratification. We briefly discuss the role of anisotropy in the occurrence and properties of such cascades. We then show that the cascade of some invariant, for example the total energy, may be transferred through nonlinear interactions to both the small scales and the large scales, with in each case a constant flux. This is in contrast to the classical picture, and we illustrate such a dual cascade in the context of atmospheric and oceanic observations, direct numerical simulations and modeling. 

We also show that this dual cascade of total energy can in fact be decomposed in some cases into separate cascades of  the kinetic and potential energies, provided the Froude and Rossby numbers are small enough. In all cases, the potential energy flux remains small, of the order of 10\% or less relative to the kinetic energy flux. 
Finally, we demonstrate that, in the small-scale inertial range, approximate equipartition between potential and kinetic modes is obtained, leading to an energy ratio close to one, 
with strong departure at large scales due to the dominant kinetic energy inverse cascade and piling-up at the lowest spatial frequency, and at small scales due to unbalanced dissipation processes, even though the Prandtl number is equal to one.
\end{abstract}   \pacs{}   \maketitle

\section{Introduction} \label{S:intro}    \subsection{The inverse cascade of kinetic energy in two-dimensional turbulence}

In 1967, Robert H. Kraichnan demonstrated, using concepts stemming from both statistical mechanics and a turbulence closure \cite{kraichnan_66}, a remarkable property of turbulent flows in two spatial dimensions (2D), namely that because of the presence of two quadratic invariants -- those preserved by truncation, namely, the energy and the integrated squared vorticity or enstrophy $Z_V=\left<|\vomega|^2 \right>, \ \vomega=\nabla \times {\bf u}$ (with ${\bf u}$ the velocity field) -- the  modal kinetic energy would migrate to the large scales of the flow in a mechanism called an inverse, or up-scale, cascade  \cite{kraichnan_67} (see also \cite{rhk_montgo, boffetta_ecke_12, pouquet_13} for reviews). This is in contrast with the generic three-dimensional (3D) case for which the energy cascades to the small scales, or down-scale. Such a vision of turbulent flows as a truncated ensemble of modes, using a decomposition of the velocity field among modes in Fourier space was  proposed by T.D. Lee \cite{lee_52} in 3D, both for fluids and magnetohydrodynamics (MHD), in the latter case predicting equipartition between kinetic and magnetic energy, in the simplest non-helical case. Such a modal energy distribution stems from the fact that, in the phase space constituted by all the modes, within the sole constraint of conservation of total energy, the solution to the Liouville equation is one of equipartition among these modes, resulting in an energy spectrum proportional to the number of modes in a given shell of radius $k$, or to $\sim k^{D-1}$, with $D$ the dimension of space.  

On the other hand, in 2D, the small-scale equipartition is that of enstrophy, the energy being transferred to the large scales in order to preserve the constancy of $Z_V$, as already foreseen in a simple case in \cite{fjortoft_53}. 
Two-dimensional flows can often be approximations to three-dimensional ones, for example in the presence of strong rotation in the atmosphere and the ocean \cite{charney_71}, in the presence of a strong imposed magnetic field, in the laboratory \cite{alemany_79} or in magnetic stars or planetary  magnetospheres, and perhaps more surprisingly in active dense bacterial systems  \cite{wensink_12}. Such inverse cascades, as diagnosed by their negative energy fluxes in the range of wavenumbers where they occur, may have been observed in the ocean \cite{arbic_13} and in the atmosphere \cite{nastrom_85, gage_86, lindborg_06} (see also \cite{waite_09} for data from numerical models).

The solutions stemming from statistical mechanics are flux-less: at equipartition, there is no more transfer between the individual modes. They have nevertheless proven to offer guidance as to what direction the energy flux takes in the forced-dissipative case in what is called the inertial range, even though the resulting energy spectrum (that is, the Fourier transform of the velocity auto-correlation function) takes a very different form, as predicted by Kolmogorov for 3D fluids \cite{K41a}. This was shown explicitly recently in direct numerical simulations (DNS) of truncated Euler flows, with a large-scale Kolmogorov spectrum and a small-scale equipartition spectrum \cite{cichowlas_05}. Because this property of energy conservation is for a system made up of a finite number of modes, $n_p$, this remains true for $n_p=3$, {\it i.e.} for each individual triadic interaction, a property called detailed balance, restricting the discussion, here, to quadratically nonlinear dynamical equations, such as is the case for incompressible fluids and MHD. The energy is transferred from scale $2\pi/k$ to scales both larger and smaller, through the convolution term resulting from the nonlinearity of the primitive equations; it is only on average that the energy is making it predominantly to the small scales in the general 3D case. It would be of interest to examine  in more detail what happens in the so-called 2D3C case, with two-dimensional geometry but three components of the  vector field, since in that case  more invariants have been identified for both neutral fluids and magnetohydrodynamic \cite{montgomery_82}. For example, for  fluids, besides total energy and total enstrophy, the extra quadratic invariant is the kinetic energy of the vertical component of the velocity which acts as a passive scalar in that case. Other fluid invariants have been discussed in the literature, such as linear or angular momentum, see e.g. \cite{dyachenko_92, newell_01, davidson_13, zakharov_15}. 

It should be noted that, in the presence of extra linear terms, the equipartition ensemble solutions are not affected theoretically. However, when  dissipation for example is added to the governing equations, a totally different distribution of energy among modes arises, but which is still consistent with an unavoidable  overall tendency of the energy to flow to the small scales. Another example is that of the interactions of turbulent eddies and waves, such as when rotation is changing the flow through the Coriolis force: in the limit of infinite rotation, the flow becomes quasi-bidimensional  (see {\it e.g.} \cite{babin_97}), and the equipartition solution is vastly different from that in a pure three dimensional case. It was shown in \cite{mininni_11} that the time to reach thermalization at a given scale grows as $\Omega^{3/4}$, where $\Omega$ is the intensity of the imposed rotation, thus resolving the apparent contradiction between the statistical mechanics prediction and the smoothing of the flow through the action of the inertial waves. This is due to long-time memory effects in such flows and leads, in the limit of infinite time, infinite Reynolds number, and infinite rotation to a lack of singularity like in the purely two-dimensional case \cite{babin_97}.

In the forced dissipative case, the inverse energy cascade in 2D exists simultaneously  with a direct enstrophy cascade to the small scales: the flux of energy is negative and constant toward scales larger than the forcing, and close to zero otherwise, whereas in the direct enstrophy cascade, the flux of enstrophy is positive and constant at scales smaller than the forcing scale, and negligible otherwise in the limit of large Reynolds numbers. This has been known for a long time, using two-point closures of turbulence \cite{pouquet_75}, direct numerical simulations (\cite{boffetta_07, boffetta_10} and references therein), and this two-cascade system was also found recently in the laboratory \cite{xia_16}.

The implied conceptual model for 2D flows was that, through nonlinear coupling between Fourier modes, the energy would reach larger and larger scales, with a constant and negative flux, and with a Kolmogorov-like spectrum, $E_V(k)\sim C_{Kol} \epsilon_V^{2/3}k^{-5/3}$, with $C_{Kol}$ a constant of order unity (but whose value is larger than in the 3D case), and with $\epsilon_V=DE_V/DT$  the total kinetic energy  dissipation rate.
On the other hand, the enstrophy  flows to small scales, again with a constant flux and with a steeper spectrum, $E_V(k)\sim k^{-3}$, both spectral indices  being obtainable by dimensional analysis through a proper choice of the invariant and of its characteristic time scale. There is likely a lack of intermittency in the inverse cascade \cite{boffetta_00}, {\it i.e.} there is no departure from dimensional scaling at all orders of structure functions, whereas such is not the case for the direct cascade in 3D. Intermittency can be diagnosed through non-Gaussian wings in Probability Distribution Functions (PDFs) of velocity gradients (and more rarely of velocities themselves \cite{paoletti_08, rorai_14}) and through a departure from self-similar scaling indicative of multi-fractality; such fat wings of PDFs correspond to localized intense structures, in the form of vortex and current sheets and filaments, as well as fronts, shocks and flux tubes.  Moreover, the two-dimensional inverse cascade for fluids has been found to be conformal invariant \cite{bernard_06}, making the bridge between turbulence and critical phenomena more evident. Conformal invariance has also been detected in three dimensions in the presence of strong rotation \cite{thalabard_11}.

The upscale transfer of energy to large scales in 2D  is also a sturdy result. For example, as shown in \cite{boffetta_07} (see also \cite{boffetta_10}), the inverse cascade does not necessitate a high Reynolds number, provided it is above unity so that nonlinearities, which are the source of the phenomenon, can counter-balance the effect of dissipation. However, the flux of energy to the large scales becomes constant only progressively, as can be observed as larger resolutions are employed; for example, in \cite{boffetta_07}, computations were performed on grids of up to $32768^2$ points (see also \cite{vallgren_11b}). A large scale separation between the forcing scale ($L_F=2\pi/k_F$ in a 2$\pi$ periodic box) and the largest scale of the flow ($L_{max}=2\pi/k_{min}$) is not really necessary, though, since a negative energy flux is already observed for a forcing wavenumber as small as $k_F/k_{min}\approx 4$, for example in the rotating stratified case \cite{aluie_11, deusebio_14b}. Note that, as the energy reaches the largest scale of the flow, an accumulation takes place, in the form of a condensation, leading to steeper spectra \cite{chertkov_07}, a topic of intense interest presently.

Even though the inverse cascade phenomenon is best observed in the presence of a forcing term, allowing for the excitation to move measurably to the large scales, it can also be detected in the decay case, as for example with a negative flux at large scale ($L>L_F$), as already found in early simulations of the decay of 2D turbulence \cite{herring_74b}. In fact, it can be observed unambiguously when performing both temporal and ensemble averages of decaying flows with  random initial conditions differing only by their phases, but with the same initial {\it rms} velocities and scale. For a given realization, the peak of the energy spectrum moves to the large scales, even though its overall amplitude, of course, diminishes with time; the envelope of that peak traces a classical $k^{-5/3}$ power law, with  the energy flux being on average negative and constant \cite{mininni_13}. This observation can be simply related to the fact that the tendency for energy to move to large scales is attributable to the nonlinearities of the flow, irrespective of the presence of a forcing term (or of a dissipative term as noted above). 

The concept of an absence of upscale transfer of energy in 3D is somewhat more fragile than what was thought some time ago; {\it a priori}, an underlying assumption is that we are dealing with an isotropic (cubic) truncation. However, in the presence of anisotropy (and anisotropic forcing), or when nonlinear interactions are limited to modes of a given helical polarity, or in the presence of inertial waves, the three-dimensional case can again lead to an inverse energy cascade as shown recently \cite{smith_96, celani_10, biferale_12, deusebio_14, herbert_14, sahoo_15, sozza_15}, with in the helical case a direct cascade of helicity, which is an invariant in the absence of dissipation and forcing  \cite{moffatt_69} (see also \cite{woltjer_60, berger_84, moffatt_92}). The possibility of an inverse cascade of energy in the presence of helicity  was already envisaged by Kraichnan \cite{kraichnan_73}, but it was ruled out in the generic case on the basis that these one-polarity interactions are swamped by all the other triadic interactions (note that a link between helicity and enstrophy is discussed in \cite{choi_09} for homogeneous isotropic turbulence, HIT).  

However, the invariance of helicity can have a profound effect on the flow. In  HIT, helicity delays the dissipation of energy, without changing the scaling laws of the energy, in either space (Fourier spectrum) or time (self-similar decay).
It has also been found that helicity can increase the Lagrangian correlation time \cite{kraichnan_77}.
In the presence of rotation or of stratification, without helicity, the temporal decay of energy is slowed down by the waves, and this can be modeled through the modification to the spatial scaling laws computed through a  weak turbulence approach (see {\it e.g.,} \cite{kimura_96, galtier_00}). When strong  helicity is present initially in the flow, this slowing-down of energy decay is again considerably weakened \cite{teitelbaum_09, rorai_13}, because of the conservation (or the quasi-conservation) of helicity. For example, in the so-called box-limited case when the integral scale cannot grow because the initial condition is of the order of the size of the computational box, the $E(t)\sim t^{-2}$ decay law for HIT becomes $\sim t^{-1}$ in  non-helical rotating or stratified flows, and $\sim t^{-1/3}$ when helicity is present  \cite{teitelbaum_09, rorai_13}. 

Finally note that, in the presence of dissipation and forcing, when there is no inverse cascade,  statistical equilibria $E(k)\sim k^{D-1}$ -- which are flux-less as noted before -- are obtained in the large scales ($L>L_F$), as shown by multiple studies over the years (for a recent study in the fluid case, see \cite{dallas_15b} and references therein). In the absence of forcing, there should be sufficient scale separation at large scale to have a functional statistical equilibrium spectrum that can lead to the 3D Kolmogorov decay law, $E_V(t)\sim t^{-10/7}$, as shown in the specific example  of 2D geometry in \cite{chasnov_97}.  

\subsection {Waves and turbulence in the context of inverse cascades}

\subsubsection{{The case of coupling to a magnetic field}}
{
Not too long after the paper of R.H. Kraichnan on the possibility of an inverse cascade of energy in 2D turbulence appeared, extensions of these ideas to systems having more than one invariant were performed, in particular in the framework of the statistical mechanics of a flow coupled to a magnetic field in the MHD framework, neglecting the displacement current in Maxwell's equations for velocities much smaller than the speed of light. Such inverse cascades in MHD exist both in two and three dimensions, for different invariants including helical ones. They have been studied using tools such as  statistical mechanics, two-point turbulence closures and direct numerical simulations, as well as in the context of the dynamics of Space Weather or of the Solar photosphere (see {\it e.g.} the recent review in \cite{pouquet_15b}).
\\
It has been shown recently that a dual constant-flux cascade to the large scales and the small scales also exists in MHD in two dimensions, for both the total energy and the ${\cal L}_2$ norm of the magnetic potential $A\hat e_z$ (with ${\bf b}=\nabla \times {\bf A}$ the magnetic field) \cite{seshasanayan_16}. Similarly, in three dimensional geometry  with only mechanical energy forcing but in the presence of a  strong imposed uniform magnetic field $B_0$, a dual cascade of kinetic energy is obtained, the amplitude of which varying with $B_0$ \cite{sujovolsky_16}. In the 2D case in particular, enough numerical resolution allows for a demonstration of the criticality of the phenomenon in terms of the control parameter which is the ratio of the magnetic to kinetic forcing amplitudes. Like for rotating stratified fluids, this may have consequences on the amount of energy transferred to the small scales of the flow, and thus on the reconnected flux available for dissipation and heating of plasmas such as the solar corona and the Solar Wind \cite{marino_08,marino_11,marino_12}.
}

\subsubsection{{Other systems with inverse cascades}}

The concept of upscale cascade has been generalized to many other physical systems, under the influence of rotation and/or stratification as in the atmosphere (\cite{charney_71, gage_79, boer_83, nastrom_85, bartello_95, augier_13} and references therein), for surface quasi-geostrophic flows, for shallow water equations \cite{warn_86}, in the Charney-Hasegawa-Mima framework \cite{bernard_07}, for capillary waves and ocean gravity waves \cite{zakharov_15}, as well as in the early universe {under the assumption of a flat space-time, leading to a phenomenon similar to a Bose-Einstein condensation which is reached in a finite time} \cite{galtier_17}.

It has also been argued that the direction of the cascade can change with physical properties of the flow, as for example for the passive scalar in the compressible case, depending on the dimension of the system and on its degree of compressibility: when the fluid becomes strongly supersonic, specifically in the sense that, in a Helmoltz decomposition, the curl-free part of the flow, ${\bf u}_C$, dominates the solenoidal div-free part, ${\bf u}_S$, then the passive scalar should undergo an upscale cascade \cite{chertkov_98}; it requires a high $u_C/u_S$ ratio, and is difficult to observe in numerical simulations of the interstellar medium at high Mach number, possibly because shocks dissipate fast, including in the case of pressure-less (infinite Mach number) Burgers turbulence \cite{pan_11}.

For stratified flows, it has been assumed that there is a decoupling between vertical layers in each of which an inverse cascade of 2D vortices could form \cite{lilly_83}, although in fact strong vertical gradients between these layers appear that destabilize such a 2D effect, as developed for example in \cite{billant_01}, leading rather to a direct cascade of energy \cite{lindborg_06}, with an important role played by the anisotropy of the flow  \cite{marino_14}.

Finally, in strongly rotating fluids in the so-called beta-plane approximation \cite{newell_69}, energy can be exchanged both ways between the sea of interacting waves and zonal flows, for Rossby wave packets, through a broadening of the resonance curve corresponding to a given dispersion relation, $\sigma(k)$. 
This idea is generalized in \cite{kuznetsov_91} where it is further conjectured  that for all such systems a link is created between coherent structures at large scales, formed because of an upscale cascade, and intermittent bursts at small scales, formed because of a direct cascade, through secondary instabilities of the large-scale structures in the form of collapsing filaments.
In the case of nonlinear optics in the framework of the nonlinear Schr\"odinger equation, the two quadratic invariants considered in 
 \cite{kuznetsov_91} are the number density and total energy. Whether these invariance properties break-down in a finite time is still unresolved, at least for 3D turbulent fluids and MHD. Similarly, one can wonder whether the break-down of large-scale structures, or condensates, in the absence of large-scale damping such as friction, leads necessarily to intermittency \cite{newell_01}.

{In the following, we consider in some detail the interactions of waves and turbulence in the presence of inverse cascades by examining these in the context of rotating stratified turbulence (RST). The equations for RST in the Boussinesq framework are given in \S \ref{S:num}, together with the dimensionless parameters and the numerical methods employed. \S \ref{S:dual} discusses first the general context of dual cascades, and then focuses on RST. Moreover, temporal and spectral data, including fluxes, are analyzed for  a set of RST runs forced at intermediate scale and already studied in \cite{pouquet_13b, marino_15p} for the relative scaling of inverse to direct energy fluxes, as observed in the atmosphere and the ocean. The role of anisotropy in the dual cascade in the context of geophysical flows is presented in \S \ref{S:aniso}, and finally \S \ref{S:discu} presents a discussion and our conclusions.}

\section{Equations and parameters for rotating stratified turbulence}   \label{S:num}    
\subsection{The Boussinesq framework}

We now write the Boussinesq equations for a stably stratified and rotating  fluid of velocity ${\bf u}$, 
vertical velocity component $w$, and density (or temperature) fluctuations around a mean gradient, $\rho$, in a physical dimension such that the ${\cal L}_2$ norm of ${\bf u}$ and $\rho$ are in units of energy per unit mass, $L^2T^{-2}$, the buoyancy itself, $b=N\rho$, having the unit of an acceleration. The rotation and stratification are assumed to be vertical, co-linear but anti-parallel, with gravity pointing downward:
\begin{eqnarray}
\frac{\partial {\bf u}}{\partial t} + \mbox{\boldmath $\omega$} \times
  {\bf u} + 2 \mbox{\boldmath $\Omega$} \times {\bf u}  &=& -N \rho \hat e_z 
  - \nabla {\cal P} + \nu \nabla^2 {\bf u} + {\bf F} \ \ , \\
  \frac{\partial\rho}{\partial t} + {\bf u} \cdot \nabla \rho &=&  Nw + \kappa \nabla^2 \rho \ , \\
  \nabla \cdot {\bf u} &=&0 \ .
\label{eq:momentum} \end{eqnarray}
${\cal P}$ is the total pressure; ${\bf F}$ is a forcing term, isotropic and random and for all three components of the velocity field, with no forcing in the thermodynamic variable, and $\hat e_z$ is the unit vector in the vertical direction. 
 Integer wave numbers correspond to a length scale for the triply periodic computational box of $2\pi$; $k_{min}=1$ and $k_{max}=n_p/3$, with $n_p$ the number of grid points in each direction, are the minimum and maximum wavenumbers of a run, with de-aliasing performed using the standard 2/3 rule. 
 
We have used the Geophysical High Order Suite for Turbulence  (GHOST) code \cite{gomez_05}, which is a pseudo-spectral code with a hybrid parallelization consisting of MPI for 1D domain decomposition, with Open MP threads for each MPI task \cite{hybrid_11}, with versatile usage from laptop to super-computers. It demonstrates linear performance scaling up to in excess of 130,000 cores. GHOST can activate the tracking of several types of Lagrangian particles, and it is easily coupled with three-dimensional visualization packages such as VAPOR \cite{clyne_07}. GHOST handles a variety of physical solvers in 2D and 3D, for fluids, MHD and Hall MHD, in both the incompressible and compressible cases. Two recent add-on features deal with  the Gross-Pitaevskii equations for superfluids. Furthermore, preliminary results indicate that GHOST implements an efficient GPU capability for the multidimensional Fourier transforms  \cite{mininni_17bb}. It also provides several small-scale modeling schemes; these models include a helical eddy viscosity \cite{baerenzung_10, baerenzung_11}, and the so-called Lagrangian alpha model 
{using a constraint on the development of small scales  through the control of a ${\cal H}_1$ norm of the velocity}
 \cite{graham_08, graham_11}.

The frequency of inertia-gravity waves for the linearized Boussinesq equations is written as:
\be 
\sigma(k)= k^{-1} \sqrt{N^2k_\perp^2+f^2k_\parallel^2} \, ,
\label{dispersion} \ee 
with $f=2\Omega$, {\it i.e.} twice the rotation frequency, and $N$, the \BV. The isotropic wavenumber is defined as $k=\sqrt{k_\perp^2 + k_\parallel^2}$, the direction $k_{\parallel}$ referring to the common direction of rotation and gravity. Geostrophic balance, with hydrostatic balance in the vertical, is obtained when neglecting all terms but the pressure gradient, the Coriolis force and the buoyancy restoring force.
In the absence of dissipation, these equations conserve the sum of the kinetic and potential energy 
$$E_T=E_V+E_P=\left<|{\bf u}|^2/2 + \rho^2/2 \right>=\int [E_V(k) + E_P(k)] dk \ , $$
also written  in terms of isotropic Fourier energy spectra of the kinetic and potential fields. Note that  kinetic helicity $H_V=\left< {\bf u} \cdot {\vomega} \right>$ is conserved only for $N=0$. In RST, it can in fact be created by the flow \cite{hide_76, hide_02, marino_13h} (see \cite{rorai_13} for the purely stratified case). Furthermore, the potential vorticity, omitting the constant $Nf$ term, is defined as:
$$
P_V= f\partial_z \rho - N\omega_z    + \vomega \cdot \nabla \rho \ ;  $$
it is a quadratic point-wise invariant of the full non-linear equations in the ideal case.

Fourier spectra are built from their axisymmetric expressions stemming from the two-point one-time velocity covariance 
as performed in  \cite{mininni_12, rosenberg_15}:
\begin{eqnarray}
e_V(|{\bf k}_{\perp}|,k_{\parallel})=
    \sum_{\substack{
          k_{\perp}\le |{\bf k}\times \hat {\bf z}| < k_{\perp}+1 \\
          k_{\parallel}\le k_z < k_{\parallel}+1}} U({\bf k}) 
  &  = \int U({\bf k}) |{\bf k}| \sin \theta d \phi = e_V(|{\bf k}|, \theta) = e_V(k, \theta) \ ,
\label{etheta} \end{eqnarray}
where  $\phi, \theta$ are respectively the longitude with respect to the $k_x$ axis and the co-latitude in Fourier 
space with respect to the vertical axis. A similar expression can be written for the potential energy density 
$e_P(|{\bf k}_{\perp}|,k_{\parallel})$ in terms of $\rho$.
The 2D-mode spectra, with no vertical variation, are  $e_{V,P}({\bf k}_\perp,k_\parallel=0)$. 
 Note that for an isotropic flow, at a given point ${\mathbf k}$ in wavenumber space, the ratio of the axisymmetric spectrum $e_V(|{\bf k}_{\perp}|,k_{\parallel})$ to the isotropic spectrum is $\sim 1/|{\bf k}|$ because the size of the volume element in the isotropic case contains an additional integrating factor of $|{\bf k}|$ compared to the axisymmetric case. Hence, if the axisymmetric spectrum behaves as $k_\perp^{-\alpha}$, then the corresponding isotropic scaling will be $k^{-\alpha+1}$. 

One can define as well the reduced perpendicular and parallel spectra \cite{sen2}
\be
E_{V,P}(k_\perp) = \Sigma_{k_\parallel} e_{V,P}({\bf k}_\perp,k_\parallel)\, , 
E_{V,P}(k_\parallel) = \Sigma_{k_\perp} e_{V,P}({\bf k}_\perp,k_\parallel)\, .
\label{ekperp} \ee
Similarly, corresponding spectra  can be written  for the helicity, namely  
$h_V({\bf k}_\perp,k_\parallel)$ and  $H_V({\bf k}_\perp)$: 

\begin{eqnarray}
h_V(|{\bf k}_{\perp}|,k_{\parallel})=
    \sum_{\substack{
          k_{\perp}\le |{\bf k}\times \hat {\bf z}| < k_{\perp}+1 \\
          k_{\parallel}\le k_z < k_{\parallel}+1}} \tilde U({\bf k}) 
  &  = \int \tilde U({\bf k}) |{\bf k}| \sin \theta d \phi = h_V(|{\bf k}|, \theta)  \ 
       =  h_\perp(|{\bf k}_{\perp}|,k_{\parallel}) + h_\parallel(|{\bf k}_{\perp}|,k_{\parallel}) \, ,
\label{htheta} \end{eqnarray}
\be
H_{V}(k_\perp) = \Sigma_{k_\parallel} h_{V}({\bf k}_\perp,k_\parallel)\,\ \  ,  \ \ 
H_{V}(k_\parallel) = \Sigma_{k_\perp} h_{V}({\bf k}_\perp,k_\parallel)\, .
\ee

 \subsection{Parameters, length scales and runs} \label{SS:param}

The fundamental dimensionless parameters for this set of equations are the Reynolds, Froude, Rossby and Prandtl numbers: 
\be
Re=\frac{U_0L_0}{\nu}, \ Fr=\frac{U_0}{L_0N}, \ Ro=\frac{U_0}{L_0f}, \ Pr=\frac{\nu}{\kappa} \ ,
\label{PARAM} \ee
with $U_0,\  L_0$  characteristic flow velocity and length scales, and $\nu=\kappa$ the kinematic viscosity and scalar diffusivity, taken equal for a unit Prandtl number. The buoyancy and rotational Reynolds numbers are defined as 
\be
R_B=ReFr^2 , \ R_\Omega=ReRo^2 \ ;
\label{RB} \ee
 $R_\omega\equiv \sqrt{R_\Omega}=\omega_{rms}/f$
{ is the micro Rossby number, 
$\omega_{rms}=[\epsilon_V/\nu]^{1/2}$ the {\it r.m.s.} vorticity, and $\epsilon_V=D_tE_V$ the effective energy dissipation rate which differs from its dimensional evaluation, $\epsilon_D=U_0^3/L_{0}$, in the presence of waves \cite{pouquet_17}. 
A micro-Froude number can be written similarly, $F_\omega=\omega_{rms}/N$, thereby linking the actual rate of energy dissipation and the \BV. Defining ${\cal R}_I=\epsilon_V/[\nu N^2]$ as the interaction parameter \cite{delavergne_16}, widely used in the analysis of mixing properties of geophysical flows \cite{ivey_08}, one has  ${\cal R}_I=F_\omega^2$, and it is easily shown that ${\cal R}_I={\cal R}_B$ only when the dissipation efficiency $\beta=\epsilon_V/\epsilon_D$ is of order one. It is found in \cite{pouquet_17}, using a simple model and analyzing numerical data, that in fact $\beta$ scales quasi-linearly with the Froude number for strongly stratified flows in the regime of wave-eddy interactions, including in the presence of weak rotation, and with a possible scaling that is shallower ($\beta\sim Fr^{0.93}$) because of rotation. This is expected from  a phenomenology of wave turbulence in which the time to transfer  energy to small scales is longer than the eddy turn-over time by a factor which is proportional to the inverse of the small parameter of the system, namely here $Fr$. 
This argument was used in \cite{marino_15p} to estimate the ratio of inverse to direct energy fluxes in RST forced at intermediate scale found to be proportional to $[Ro Fr]^{-1}$(see equ. (\ref{RPI})), with a role for both the Froude and the Rossby numbers, and similarly to  justify the observed scaling for the mixing efficiency $\Gamma_f$ in RST, {\it e.g.} $\Gamma_f \sim Fr^{-1}\sim {\cal R}_B^{-1/2}$, in the high $Re$ turbulence regime of such flows and for $Fr>0.2$.} 

Finally, the following length-scales can be monitored:
\begin{eqnarray}
L_{int} &=& 2\pi E_V^{-1}\int [E_V(k)/k]dk \ , \ \  L_B=2\pi U_0/N \ , \ \  L_{Ell}=2\pi \sqrt{E_P}/N, \  
 \label{length1}  \\
\ell_{Oz} &=& 2\pi [\epsilon_V/N^3]^{1/2} \ , \  \  \ell_{Ze}=  2\pi [\epsilon_V/f^3]^{1/2} \ ,  \ \  \eta= 2\pi [\epsilon_V/\nu^3]^{-1/4}\ ;
\label{length2} \end{eqnarray}
these are the integral scale, the buoyancy and Ellison scales $L_{B,Ell}$, the Ozmidov and Zeman scales, $\ell_{Oz, Ze}$, at which isotropy presumably recovers, and the dissipation scale $\eta$, based on an isotropic Kolmogorov energy spectrum in the small scales. Note that the Zeman scale was named the Hopfinger scale by Carl Gibson \cite{gibson_91}.
The Rossby deformation radius is $L_D=[N/f]L_\parallel$, and $\epsilon_V=\nu Z_V$; $\epsilon_P=\kappa Z_P$ are the kinetic and potential energy dissipation rates, with $Z_V=\left< |\nabla \times {\bf u}|^2 \right>$ (already defined), and 
$Z_P=\left< |\nabla \rho|^2 \right>$. Note that $L_B/\ell_{Oz}\sim Fr^{-1/2}$, and that $\ell_{Oz}/\eta \sim {\cal R}_B^{3/4}$, showing that the buoyancy Reynolds number in stratified flows plays an equivalent role to the Reynolds number in homogeneous isotropic turbulence, when the Ozmidov scale is taken instead of the integral scale, since the isotropic range of the stratified turbulence starts at $\ell_{Oz}$, with presumably a Kolmogorov $k^{-5/3}$ spectrum. In strongly stratified turbulence, care must be taken in the evaluation of the energy dissipation rate on which the Ozmidov scale is based since, as stated before, $\beta=\epsilon_V/\epsilon_D$ 
now differs from unity (see \cite{pouquet_17} for more details).

In order to measure the small-scale and large-scale anisotropy of the flow, one can also define the  perpendicular and parallel  integral and Taylor scales as:
\be 
L_{\perp,\parallel} = 2\pi \frac{\Sigma k^{-1}_{\perp,\parallel} E_V(k_{\perp,\parallel})}{\Sigma E_V(k_{\perp,\parallel})} \ , \ \ 
\lambda_{\perp,\parallel} = 2\pi \  \large[ \frac{\Sigma E_V(k_{\perp,\parallel})}{\Sigma k^{2}_{\perp,\parallel} E_V(k_{\perp,\parallel})} \large]^{1/2}  \ , 
\label{length3} \ee
where $\Sigma$ stands for a sum from either $k_\perp=1$ or $k_\parallel=1$ to $k=k_{max}$.

The buoyancy scale corresponds to the thickness of the layers in the purely stratified case, and to the scale at which the ratio of kinetic to potential energy reaches its minimum in RST \cite{marino_15w}. Finally, the Ellison scale is associated with the vertical distance traveled by a fluid particle before being completely mixed \cite{ellison_57} (see e.g. \cite{barry_01, smyth_01, chung_12, mater_14b}). The Thorpe scale \cite{thorpe_77, metais_89, herring_89, thorpe_12b} was introduced to estimate the length scale at which turbulent over-turning occurs in stratified flows to recover stability; it was found to be comparable to the Ozmidov scale at which isotropy is recovered in \cite{dillon_82}, whereas it was found to be proportional to the Ellison scale  in a series of recent numerical simulations \cite{waite_06b, mater_13}.

\section{Dual cascades of energy in the three dimensional case}  \label{S:dual}
\subsection{General context}

A new paradigm has emerged recently, that of a \emph{dual cascade}, by which it is meant that the \emph{same} invariant (say, total energy) undergoes a self-similar cascade \emph{both} to the large scales and to the small scales, \emph{in both cases with constant fluxes, of opposite signs and of different magnitudes {\it a priori}},  the ratio of the direct to inverse flux being an unknown of the problem. We now briefly review some of these systems in which a dual cascade is observed. 

{To our knowledge, the first indication of such a dual cascade can be found in \cite{smith_96}. Because the numerical resolution at the time was limited, several features were implemented to achieve this, perhaps somewhat unexpected, result. First, a small aspect ratio of the computational box was used, in order to enforce as is possible the two-dimensionality of such a flow; this further allowed for a sufficient scale separation in two inertial ranges, at scales smaller and larger than the forcing scale. Moreover, both an eddy viscosity model and an hyper-diffusivity (with a higher power of the Laplacian operator) were used in some cases. Finally, the forcing was two-dimensional and with only two components (2D2C). This allowed for an explicit demonstration of the coexistence of an inverse and direct cascade and, as pointed out by the authors, it included a physical behavior which was more general than quasi-geostrophy, which contains only two-dimensional dynamics with a (sole) inverse cascade of energy, causing a problem of interpretation of atmospheric and oceanic data which show clear and strong mixing at small scales (see {\it e.g.} \cite{peltier_03, mcwilliams_16} for reviews). This approach was further analyzed, with substantially more resolution, in \cite{celani_10}. The ``splitting'' of turbulent cascades, for the passive scalar variance for compressible flows, and for the kinetic energy, was associated with the  link between Eulerian statistics and Lagrangian particles. Under basically the same conditions as in \cite{smith_96},  the dual cascade was shown to depend on the ratio of the vertical length scale $L_z$ to the forcing scale $\ell_f$; it can then be interpreted as a phase transition with control parameter $L_z/\ell_f$.}

When performing  both laboratory experiments and numerical simulations on surface capillary waves in superfluid helium, it is found in \cite{abdurakhimov_15} (see also \cite{kolmakov_14}) that again a  dual energy cascade occurs, this time in the absence of a second invariant such as enstrophy. In this case, the dual cascade is attributed to providing a means of replacing large scale energy that is lost due to the existence of a large-scale dissipation process; this allows for replenishing the large-scale energy dissipated by friction. 
The formation of these large-scale coherent structures can be diagnosed through the wings of the Probability Distribution Function of the wave amplitude, {\it i.e.} to a departure from the Gaussianity assumed  in weak turbulence. Thus, the dual cascade provides a link between large and small scales,
 {but with cascades retaining their locality in scale interactions (with spectral laws between $k^{-1}$ and $k^{-3}$).}
  The energy flux towards large scales (here, in terms of temporal frequencies) is seen to increase as the damping coefficient is increased, whereas the flux to small scales, which is not shown, is well-known to exist corresponding to the traditional Kolmogorov-Zakharov spectral law for these waves in the weak turbulence regime. 

An inverse energy cascade was observed in experiments \cite{ganshin_08}  on second sound waves in superfluid Helium above a critical threshold in the  amplitude of the driving, with the generation of sub-harmonics, {\it e.g.}, through a modulational instability as also observed for rogue waves in the ocean \cite{dysthe_08}. The ``sharing'' of the energy flux between the large and small scales corresponds to a reduction of the energy being transferred to the small scales, and results in a lesser amount of dissipation and mixing, as reviewed in \cite{ivey_08} (see also, e.g., \cite{mater_14} for stratified flows, 
{and \cite{marino_15p} for RST).}

Finally, the presence of a dual cascade is also hinted at in the numerical data of \cite{boffetta_12} for the Rayleigh-Taylor instability, with a small-scale Kolmogorov cascade and a Bolgiano-Obukhov (BO) scaling in the large scales for the energy spectrum, identified through a reversal of sign of the third-order structure function corresponding to the nonlinear energy flux, the small-scale flux being called a ``remnant''. It is also noted that the vertical buoyancy flux is large in the large scales (see \cite{kumar_14, rosenberg_15, verma_17} for the unforced rotating stratified case for which BO scaling is also identified). The resultant dual energy cascade is attributed to the presence of a transitional regime between 2D and 3D behavior, 
{as clearly demonstrated in MHD in two and three space dimensions \cite{seshasanayan_16, sujovolsky_16}.}

\subsection{Temporal evolution of rotating stratified turbulence}

 \begin{table}  \caption{   \label{tab_Y} 
 DNS parameters of the runs discussed in this paper. Id is an identification, a star indicating runs on grids of $2048^3$ points, the other runs being on grids of $1024^3$ points. Run 1 also has $k_F=7.5$, in order to maximize the Reynolds number; all other runs have $k_F=10.5$, and $2\pi/k_F$ is taken as the length scale entering the expression of the dimensionless parameters. $Fr$ and $N/f$ are the Froude number and the ratio of the \BV\ to twice the rotation frequency. $Re, \ {\cal R}_B$, $Ro$ and $R_\omega$ are the Reynolds and buoyancy Reynolds numbers, the Rossby and micro Rossby numbers. $R_\Pi$ defined in equ. (\ref{RPI}) is the ratio of inverse to direct kinetic energy fluxes.
 { Finally, $\epsilon_V$ and $\beta=\epsilon_V/\epsilon_D$ are the kinetic energy dissipation rate and  its  adimensionalized value, with $\epsilon_D=U_0^3/L_0$, measured at the time $t_M$ of peak of dissipation for each run.}
 C123, C45 and C6 are the color codes used for these runs in Figs. \ref{f:april-Ta}--\ref{f:april-Tc}, Figs. \ref{f:april-Sa}, \ref{f:april-Sb}, and  in Fig. \ref{f:april-Fa}. All these runs show an inverse flux of potential energy (see Fig. \ref{f:april-Fa}), with a measurable energy growth (see Fig. \ref{f:april-Tb}).   
 {$1/[Ro Fr]$ is the parameter identified in \cite{marino_15p} for establishing the scaling of $R_\Pi$.}
  } 
 \begin{tabular}{cccccccccccccccccc} 
 Id  & $\ Fr$ & $\ N/f\ $ &   Re   & ${\cal R}_B \ $ & $\ Ro \ $& $R_\omega$ &  $R_\Pi \ $ & $t_M$ & $\epsilon_V$ & $\beta$ &  C123 & C45 & C6 & $[Ro Fr]^{-1}$  \\ 
\hline \hline
1* $^+$ & .028 & 5. & 38900 & 30 & .14 & 27 & 11. & 2.8 & .06 \ & $4.5 \ 10^{-3}$ & black & - & - & 258 \\
2* $ $ & .047 & 7. & 19800 & 44 & .33 & 47 & 3.6 & 3.6 & .16 \ & $2.4 \ 10^{-2}$ & red & - & - & 64 \\
\hline
3 $ $ & .061 & 8. & 7560 & 28 & .48 & 42 & 1. & 2.7 & .16 \ & $3.3 \ 10^{-2}$ & blue & cyan & blue & 34 \\
4 $ $ & .063 & 5. & 7850 & 31 & .31 & 28 & 2.2 & 2. & .12 \ & $2.4 \ 10^{-2}$ & magenta & black & green & 50 \\
5 $ $ & .073 & 10.5 & 7270 & 39 & .77 & 65 & .3 & 2.9 & .19 \ & $5.2 \ 10^{-2}$ & green & blue & cyan & 18 \\
6 $ $ & .088 & 5. & 7320 & 57 & .44 & 38 & 1. & 2. & .19 \ & $6.6 \ 10^{-2}$ & cyan & green & black & 26 \\
7 $ $ & .089 & 8. & 6360 & 51 & .71 & 57 & .2 & 1.5 & .16 \ & $4.2 \ 10^{-2}$ & dash & magenta & magenta & 16 \\
\hline    \end{tabular}  \end{table}  

In rotating stratified flows, two specific characteristics are that they carry dispersive inertia-gravity waves, and that they are anisotropic. This implies that, for some parameter regimes, the dynamics will be governed by the interplay of waves and eddies, at least for a range of scales in which the ratio of the nonlinear frequency to the wave frequency is small, as for small Rossby or Froude numbers.  
Taking for $f$ the value of $10^{-4}\ s^{-1}$ corresponding to mid latitudes, $N/f$ is typically $\approx 100$ in the atmosphere, and $\approx 10$ or less in the ocean. 

A dual energy cascade was already found in \cite{mininni_10a} in the context of purely rotating flows in the presence of helicity. In that case, the energy flux is constant at scales both larger and smaller than the forcing scale, with opposite signs, whereas the helicity only goes to the small scales. Furthermore, it was concluded that the helicity flux, normalized by the scale of the forcing, becomes progressively dominant over the flux of energy to small scales as the rotation increases in intensity, recovering a single energy cascade dynamics in the limit of infinite rotation.

The switch from a wave-dominated to a vortex-dominated regime in such flows was studied for example in \cite{sukhatme_08} as a function of the Burger number, defined as $N^2/f^2$, with the large scales dominated by the slow eddy mode and the small scales by the wave modes (see also \cite{kurien_12}).
In flows dominated by stratification (large $N/f$) as in the atmosphere, the small scales are characterized mostly  by the wave modes, whereas the situation is more complex for the opposite case of rotation dominance, with a transitional wavenumber.
This double cascade, including in regimes far from geostrophic balance, can be interpreted, as done in \cite{bartello_95}, as an inverse cascade of the purely 2D geostrophic (eddy) mode, and a direct cascade of the ageostrophic field interacting with the former in a catalytic way.

Such a dual energy cascade can be an essential feature of the overall dynamics of the ocean, and of the atmosphere, providing a bridge between the large scales in quasi-geostrophic equilibrium due to a balance between the Coriolis force, gravity and the pressure gradient, and undergoing an inverse energy cascade, and the small scales likely fed by a direct energy cascade allowing for the dissipation of the energy injected for example through solar radiation, wind forcing on the surface of the ocean and tidal waves (see \cite{mcwilliams_16}  for a recent review). 
The dual cascade could be attributed to the fact that, in a wave-vortex decomposition, the slow mode and the wave modes are not constrained in the same way by invariance properties, with a degeneracy of the quadratic  invariant of potential enstrophy (in the quasi-linear limit)  \cite{herbert_16}.
 In fact, for purely rotating flows, it was shown in \cite{chen_05} that  the inverse energy cascade in that case could be attributed  to the slow modes with zero frequency, {\it i.e.} the modes with $k_\parallel=0$.
 
\begin{figure*}   
\includegraphics[width=.77\textwidth]{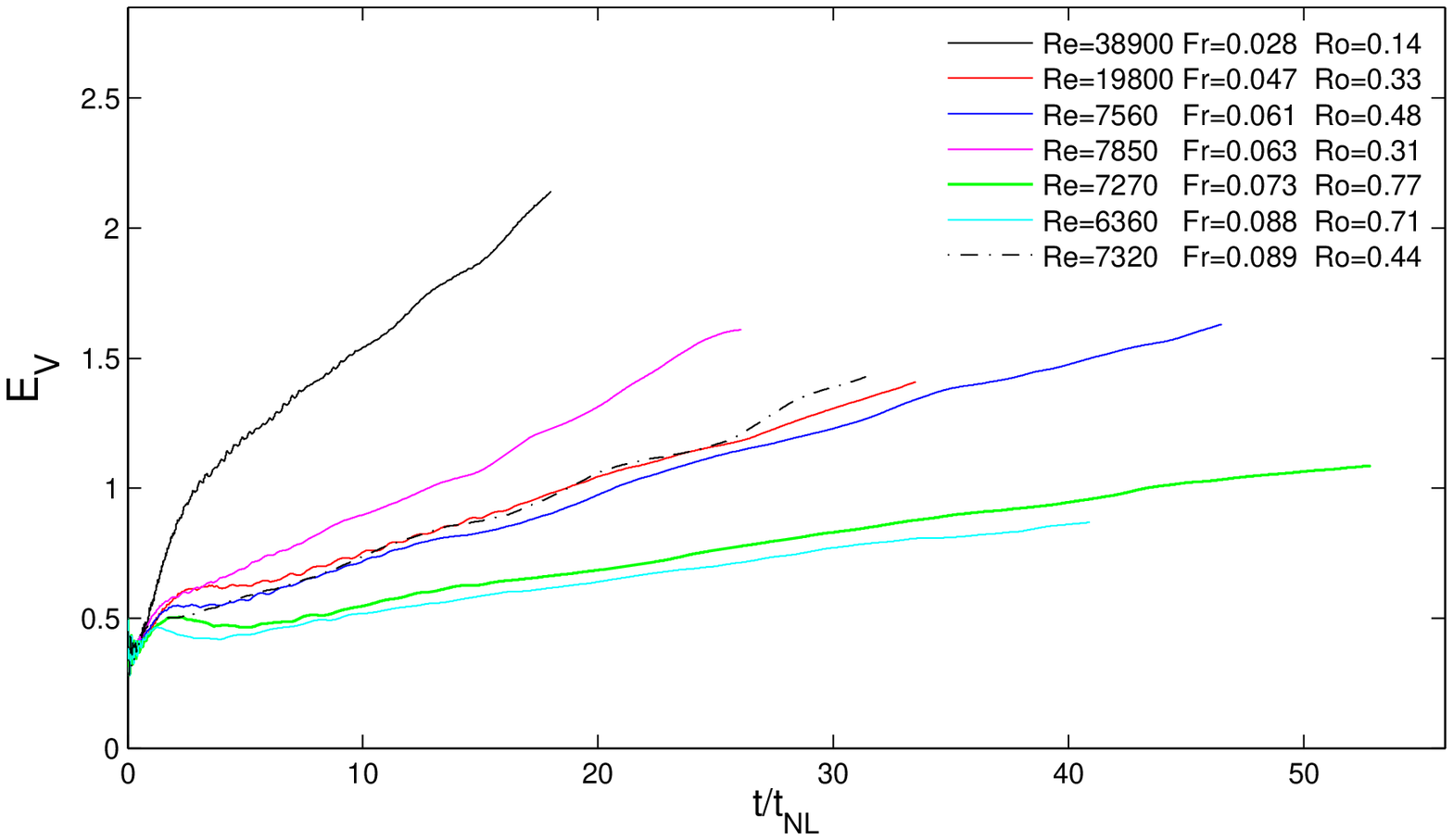}    
\includegraphics[width=.77\textwidth]{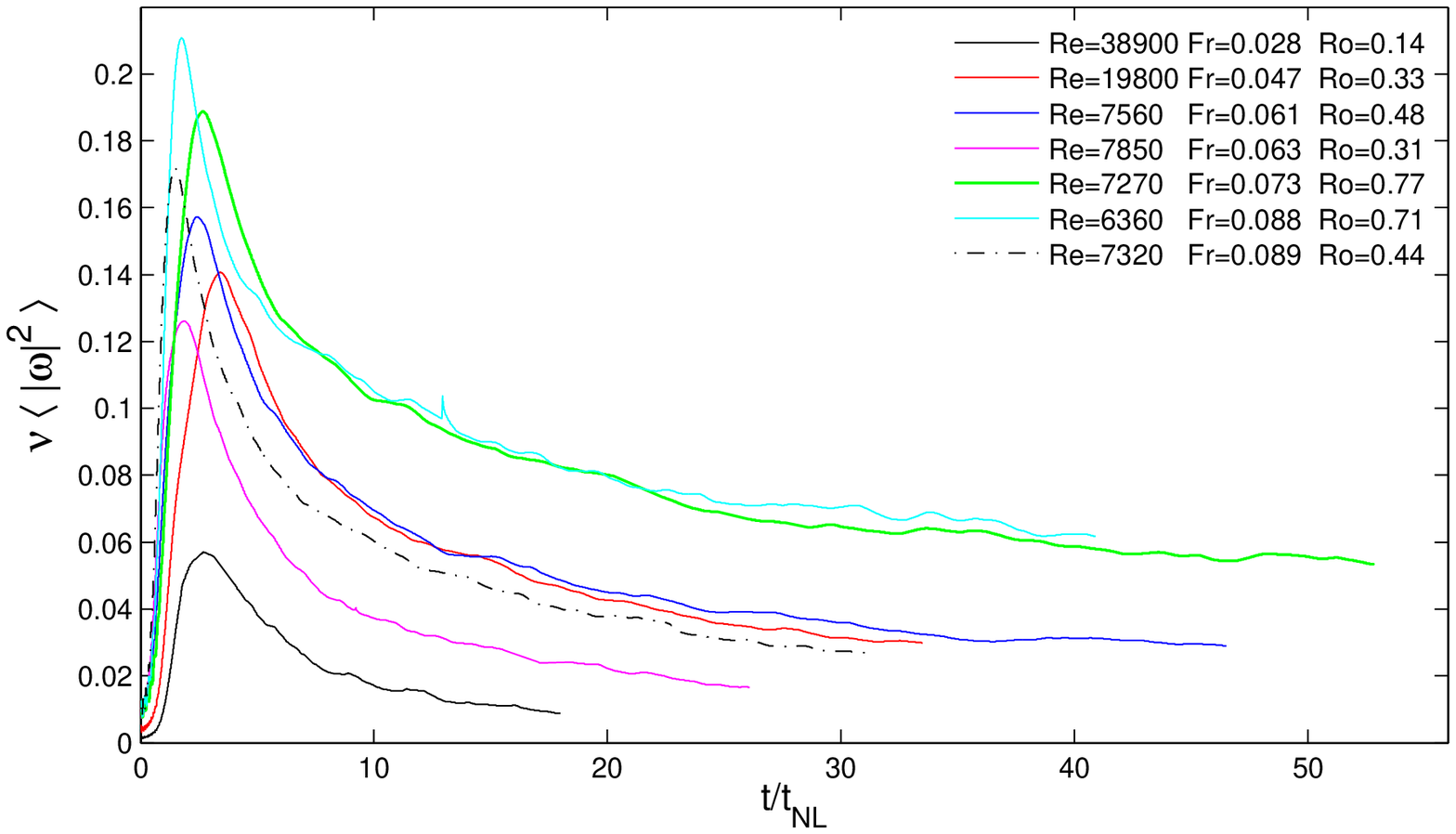}   
\caption{Temporal evolution, in units of turn-over time, of the kinetic energy $E_V$ (top) 
and of its  dissipation rate (bottom), $\epsilon_V = \nu \langle | \nabla \times {\bf v}|^{2} \rangle = \nu Z_V$; the plots are color-coded for 
different parameters (see inserts and Table \ref{tab_Y}). } 
\label{f:april-Ta} \end{figure*}

The inverse cascade  is less efficient as the Rossby number is increased. For example, in Fig. \ref{f:april-Ta} (top) is shown the temporal evolution of the kinetic energy for various  values of $N/f$, i.e. for different relative intensities of the imposed rotation and stratification, for the runs listed in Table \ref{tab_Y}; these runs were succinctly analyzed in \cite{marino_15p} to study the scaling with parameters of the direct and inverse fluxes and of their ratio. The two runs at high Reynolds numbers were done on grids of $2048^3$ points, the other runs on grids of $1024^3$ points, and the unit of time is the turn-over time  $\tau_{NL}=L_0/U_0$, computed for each flow. 

 The fastest growth is for the lower Rossby number, and the growth rate varies with Rossby number as well as with $N/f$ \cite{marino_13i}, where  it is also shown that the stronger growth rate occurs when there are no resonances, that is for $1/2 \le N/f \le 2$ \cite{smith_02}; no growth is found in the purely stratified case. At the bottom in Fig. \ref{f:april-Ta} is given the kinetic energy dissipation for the same runs.
Note the substantial growth of $E_V$ and  the corresponding smaller growth 
of $\epsilon_V = \nu Z_V$ for the runs with strong rotation 
{(runs 1, 2, and 4)}: the more energy goes to the large scales in an inverse cascade, the less energy goes to the small scales, where it is available for dissipation. All runs reach a peak of dissipation within a few turn-over times ($t/\tau_{NL}$ 
{between $\approx 1.5$ and $\approx 3.6$}), with a slight dominance of kinetic energy over potential energy,  immaterial of the rotating and stratified parameters. Also, a possible saturation of dissipation may be reached at long times. Here, $N/f$ varies by roughly a factor of 2, and the main effect is  that of the Rossby number. It was shown in \cite{marino_13i} that the rate of increase of the kinetic energy depends on $N/f$ as well (see also \cite{pouquet_13}), and the only cases showing a decrease -- after an initial transient where the dynamics is dominated by kinetic-potential energy exchanges induced by inertia-gravity waves -- are cases with no rotation ($f=0$) \cite{marino_13i}.

\begin{figure*}   
\includegraphics[width=.77\textwidth]{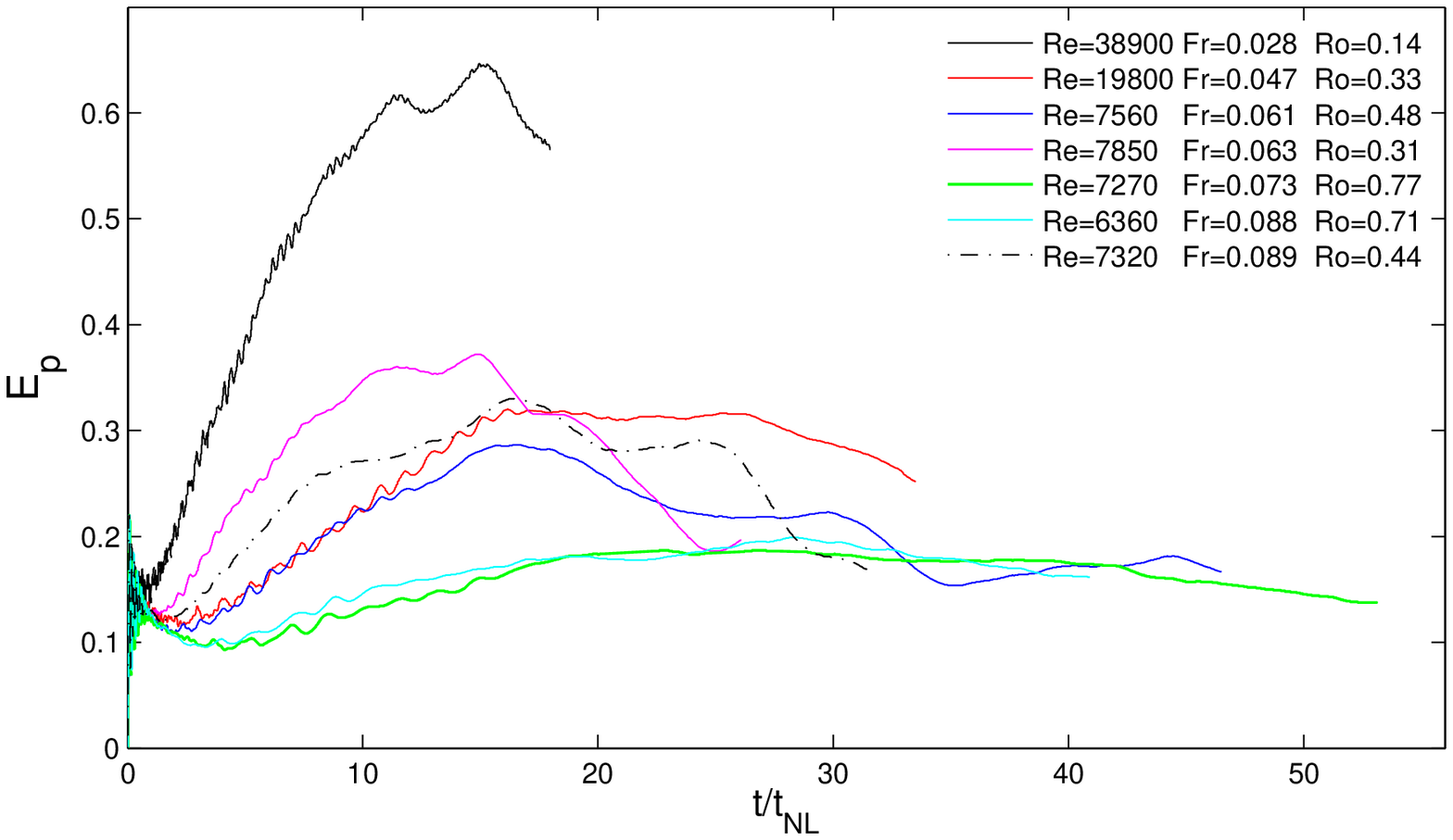}   
\includegraphics[width=.77\textwidth]{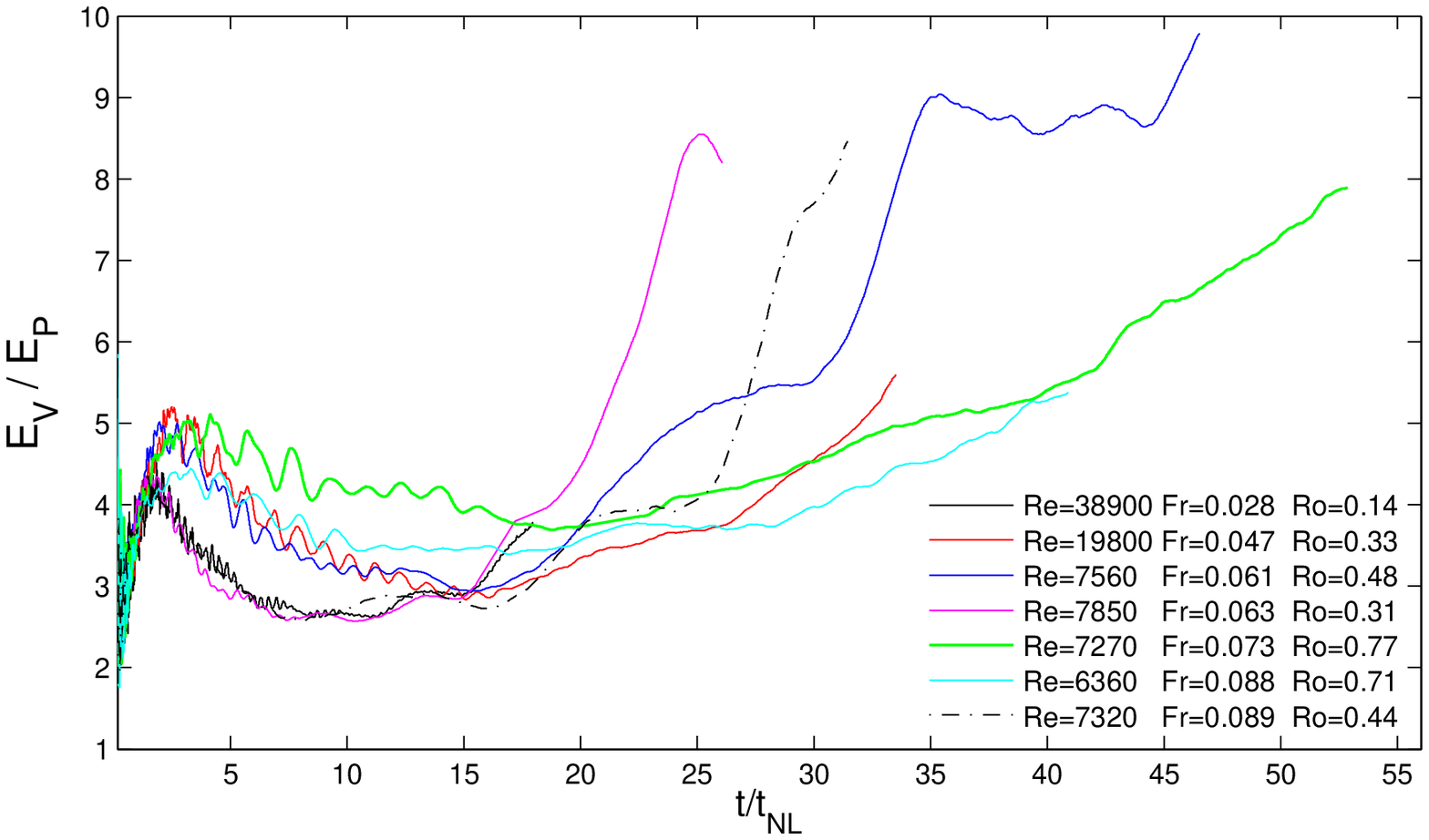}  
\caption{Temporal evolution, with $t_{NL}$ the turn-over time, of the potential 
energy (top), and of the ratio of kinetic to potential energy $E_V/E_P$ (bottom), 
for the same runs as in Fig. \ref{f:april-Ta}.
} \label{f:april-Tb} \end{figure*}

The runs considered in this paper have been selected for the fact that potential energy is also observed to grow measurably with time, as seen in Fig. \ref{f:april-Tb}, which displays the temporal evolution of the potential energy  (top) and of the ratio of kinetic to potential energy, $r_E=E_V/E_P$
 (bottom), for the same runs. Again, the rate of growth of $E_P$ is larger the stronger the rotation. The growth of $E_P$ is followed by a rough plateau with, for most runs, a slowly decreasing trend. This plateau is reached at later times for the runs with  the two largest Froude numbers. 

The evolution of the ratio of kinetic to potential energy, $r_E$ (Fig. \ref{f:april-Tb}, bottom) is more complex. It displays several temporal regimes. First, a strong growth occurs which corresponds to the flow developing under the forcing from zero initial conditions and up to the peak of dissipation. At that time, there is a high ($\approx 5$) relative kinetic  energy.  A second phase follows in which the waves are influencing the dynamics, drawing the kinetic and potential energy towards equipartition, but not quite.  But as the kinetic energy grows, the Rossby number increases and a third temporal phase begins, with a strong relative increase of kinetic energy that takes over, with no clear saturation for the runs, after up to 50 turn-over times, or more than 500 $N^{-1}$. The starting time of that third phase depends inversely on Rossby number, and after that time waves are almost not discernible any longer for the evolution of these global quantities. This is consistent with the fact that the growth of potential energy decreases when the rotation weakens, and the peak of the growth is delayed, whereas no such slowing-down of the growth of kinetic energy is observed, the smallest wavenumber of the system having been barely reached at the end of these three-dimensional computations, so that no sizable finite-size box-limited effect are felt yet, although some is visible in the energy spectra as a function of $k_\perp$. 

\begin{figure*}   
\includegraphics[width=.77\textwidth]{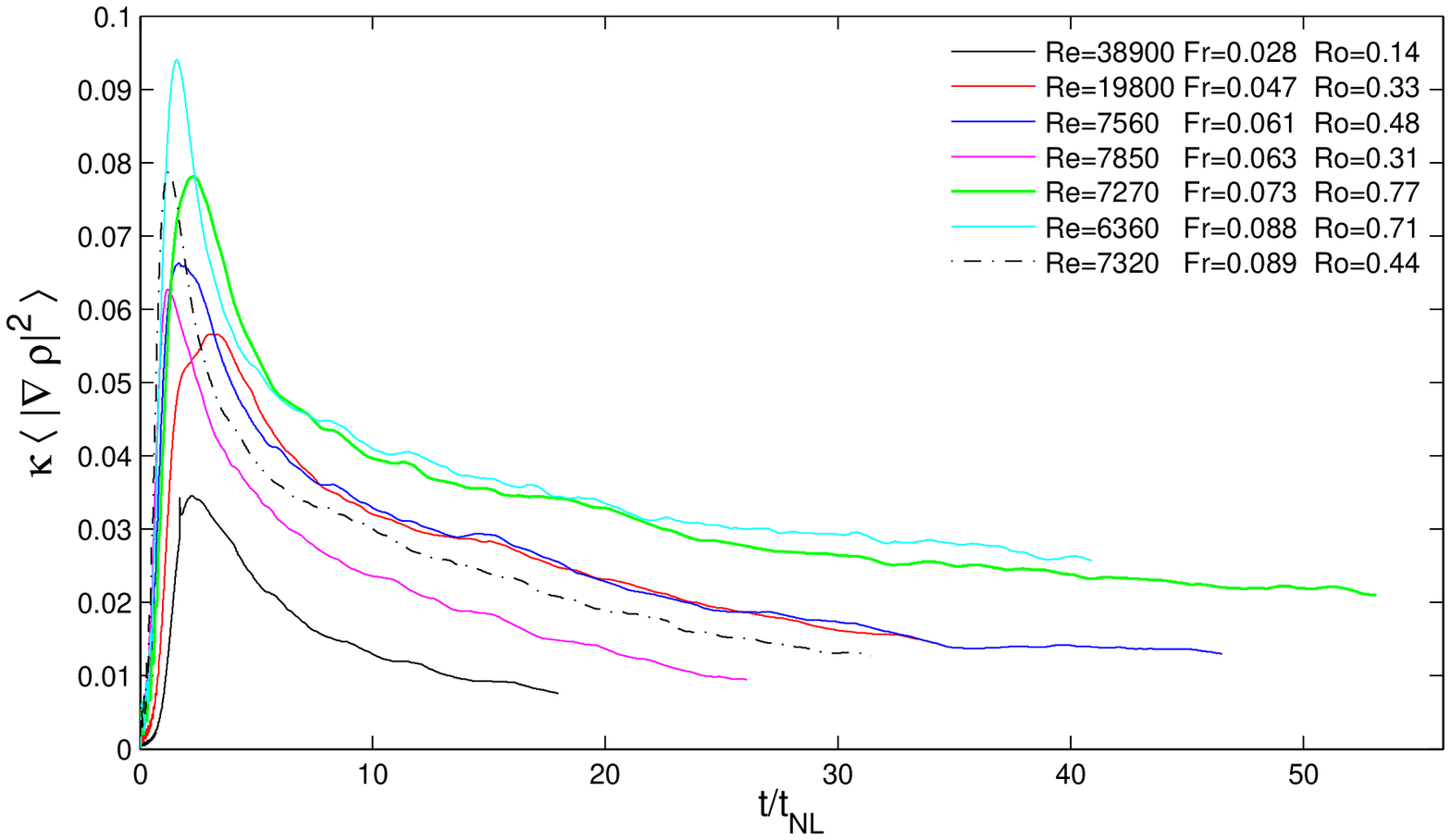}   
\includegraphics[width=.77\textwidth]{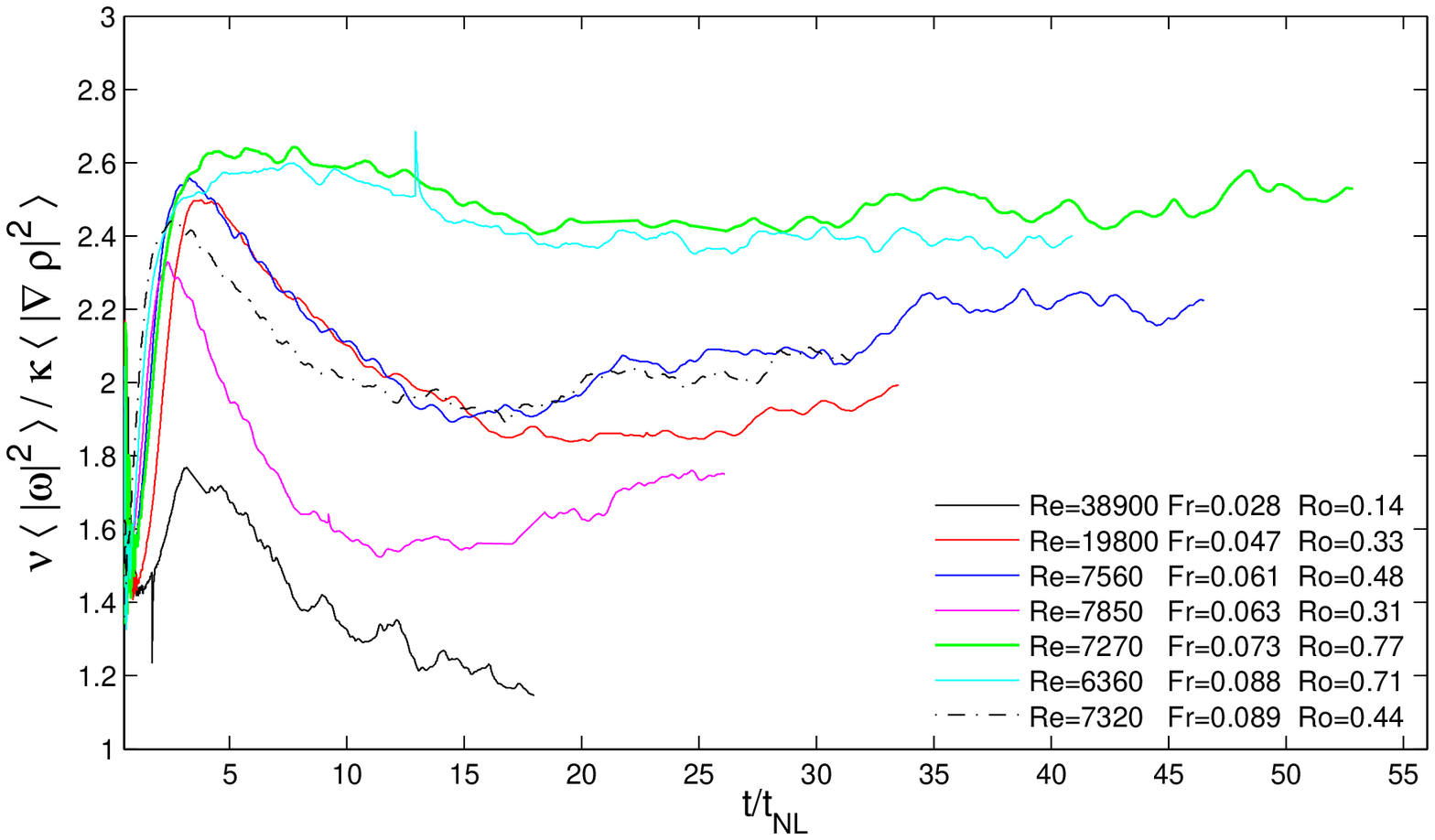}  
\caption{Temporal evolution, as a function of  $t/t_{NL}$, of the potential energy dissipation $\epsilon_P = \kappa \left< |\nabla \rho|^2 \right> = \kappa Z_P$ (top), and of the ratio of kinetic to potential energy dissipation $Z_V/Z_P$ (bottom), for the same runs as in Fig. \ref{f:april-Ta}.
} \label{f:april-Tc} \end{figure*}

Finally, in Fig. \ref{f:april-Tc}, we show the potential energy dissipation $\epsilon_P = \kappa \left< |\nabla \rho|^2 \right> = \kappa Z_P$ as a function of time (top) and the ratio of kinetic to potential energy dissipative rates, $Z_V/Z_P$ (bottom), using $\nu = \kappa$. 
With powerful inverse cascades at low Rossby numbers, part of the energy available for dissipation is in fact stored in the large scales, and moreover, with small  Froude numbers as here, one is in the regime of stratified turbulence at intermediate buoyancy Reynolds numbers (see {\it e.g.} \cite{mater_14, rosenberg_16, pouquet_17}).
The potential energy dissipation is similar to its kinetic counterpart, although, when examining their ratio, there is again a clear trend in Rossby number, with more kinetic energy dissipation as the rotation weakens:
the three runs with the lowest Rossby numbers have the lowest $E_P/E_V$ ratio, as expected, leading to the lowest peak of both kinetic and potential dissipation for these flows as well (see Figs. \ref{f:april-Ta}-\ref{f:april-Tb}). 
{This may have an impact on the scaling with parameters of the mixing efficiency of RST, as it is sometimes evaluated on the ratio of potential to kinetic energy dissipation \cite{mater_14, pouquet_17}.
}

 \subsection{Spectral data}

\begin{figure}[ht]    
                   \includegraphics[width=.45\textwidth, height=14.1pc]{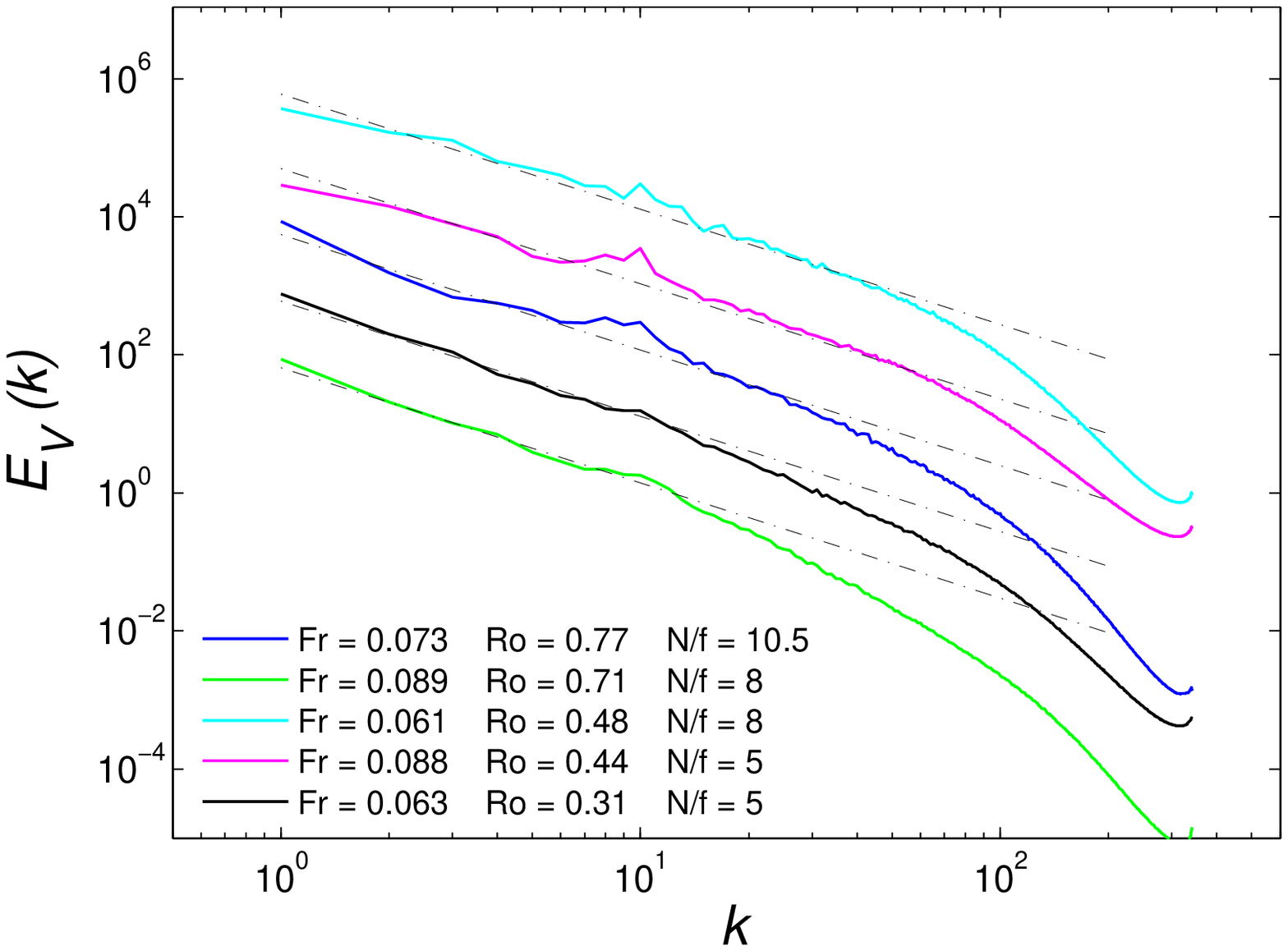}  
                   \includegraphics[width=.45\textwidth, height=14.1pc]{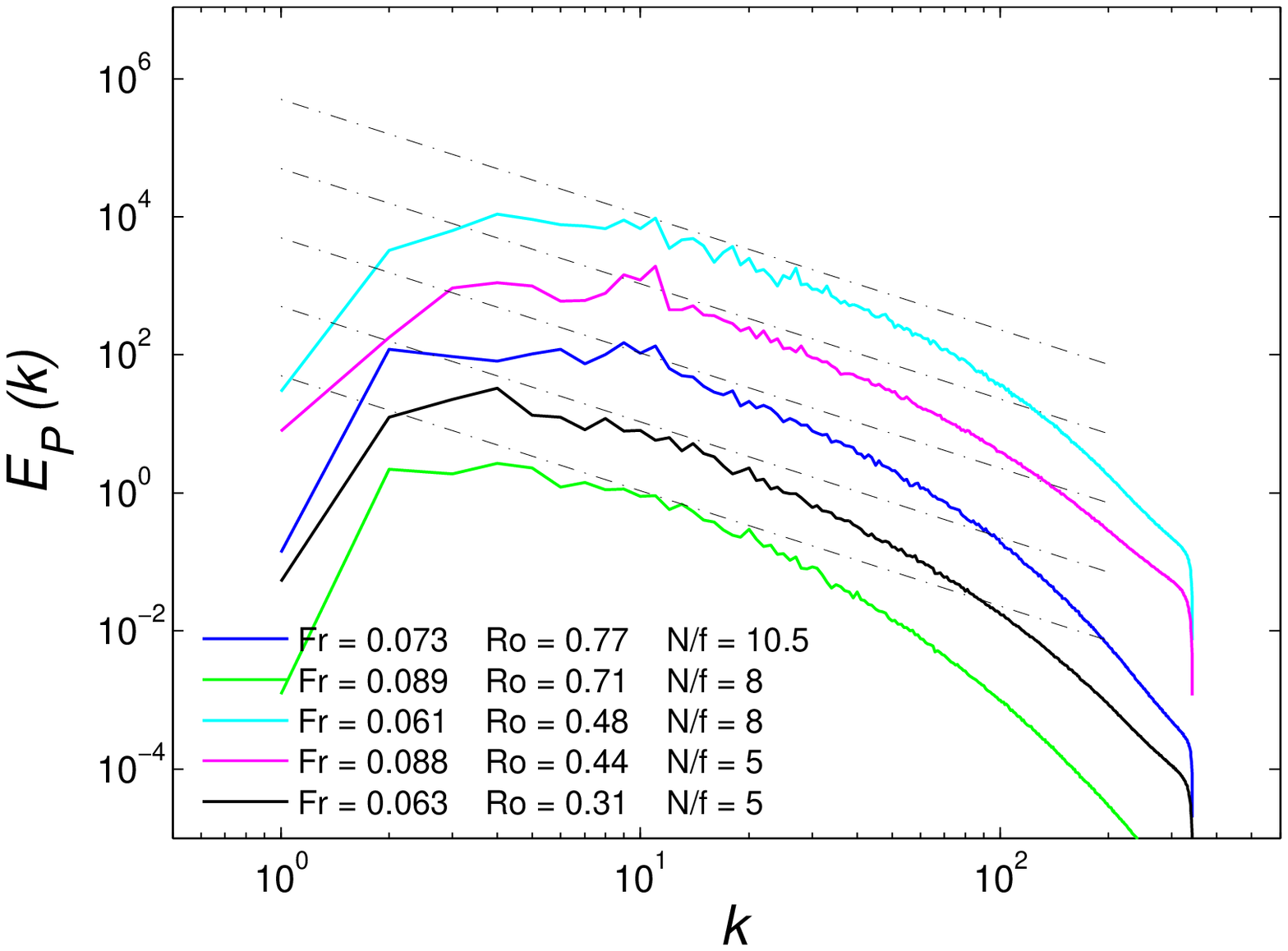}     
                   \includegraphics[width=.45\textwidth, height=14.1pc] {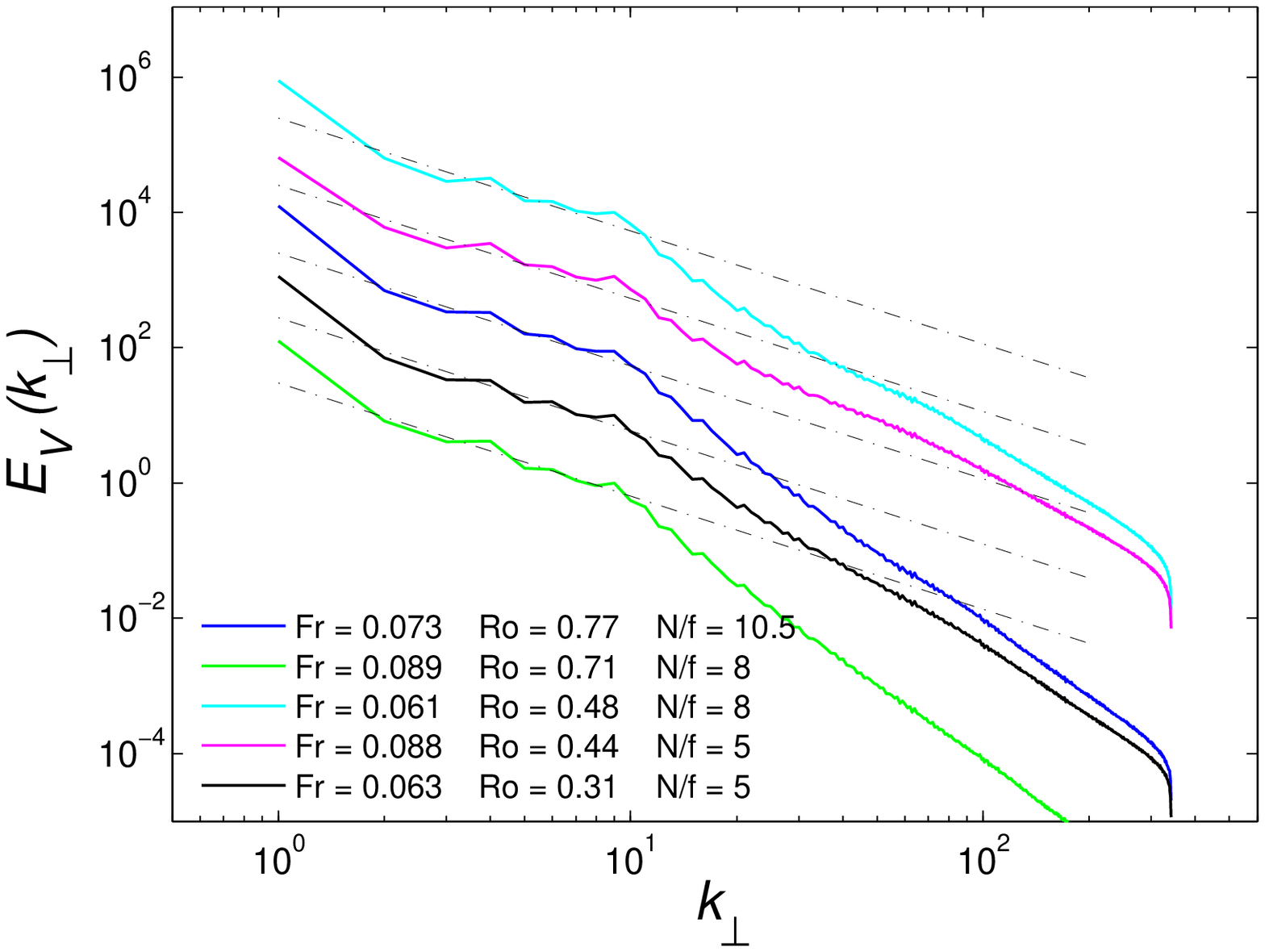}      
                   \includegraphics[width=.45\textwidth, height=14.1pc]{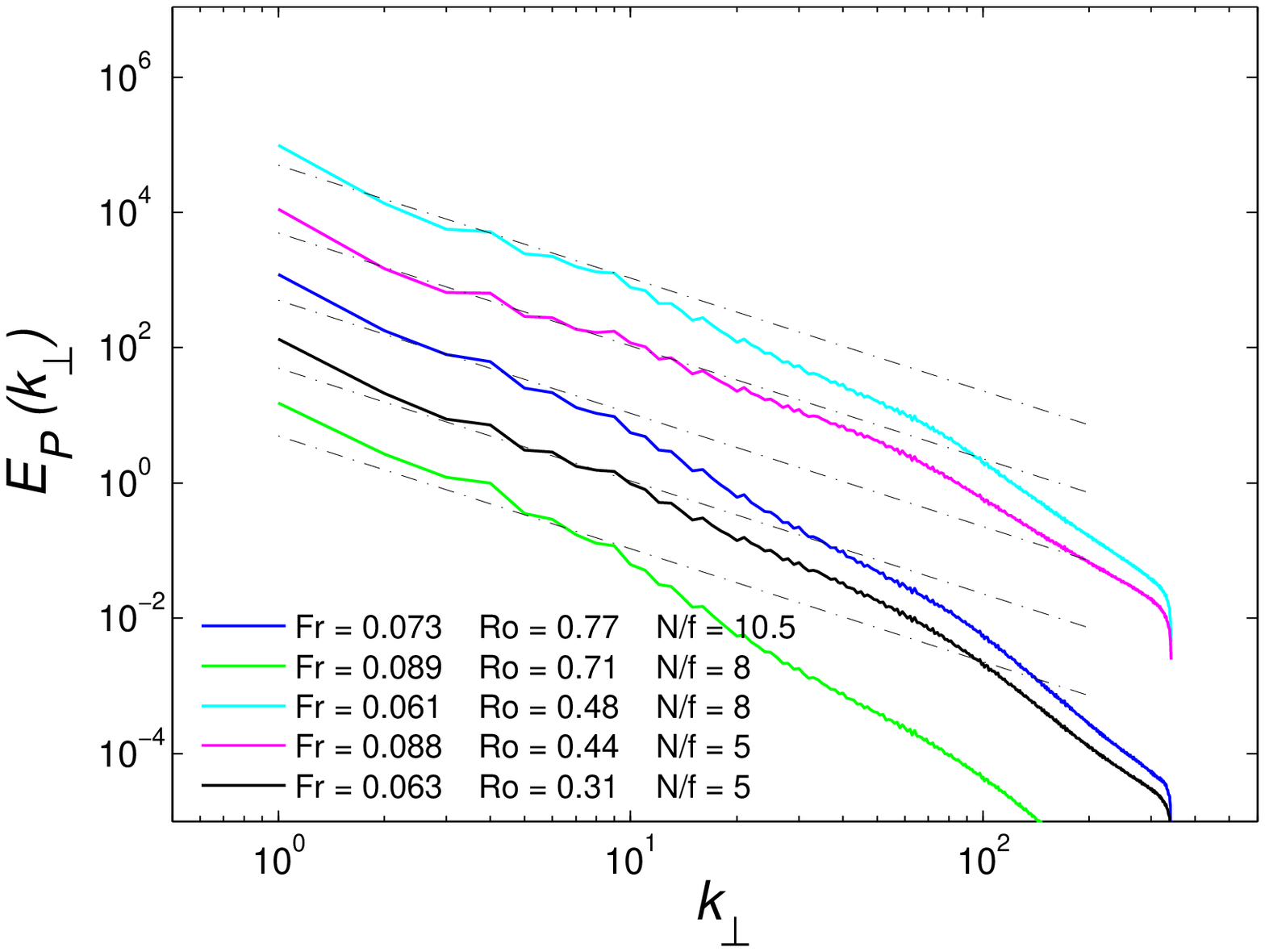}     
                    \includegraphics[width=.45\textwidth,height=14.1pc] {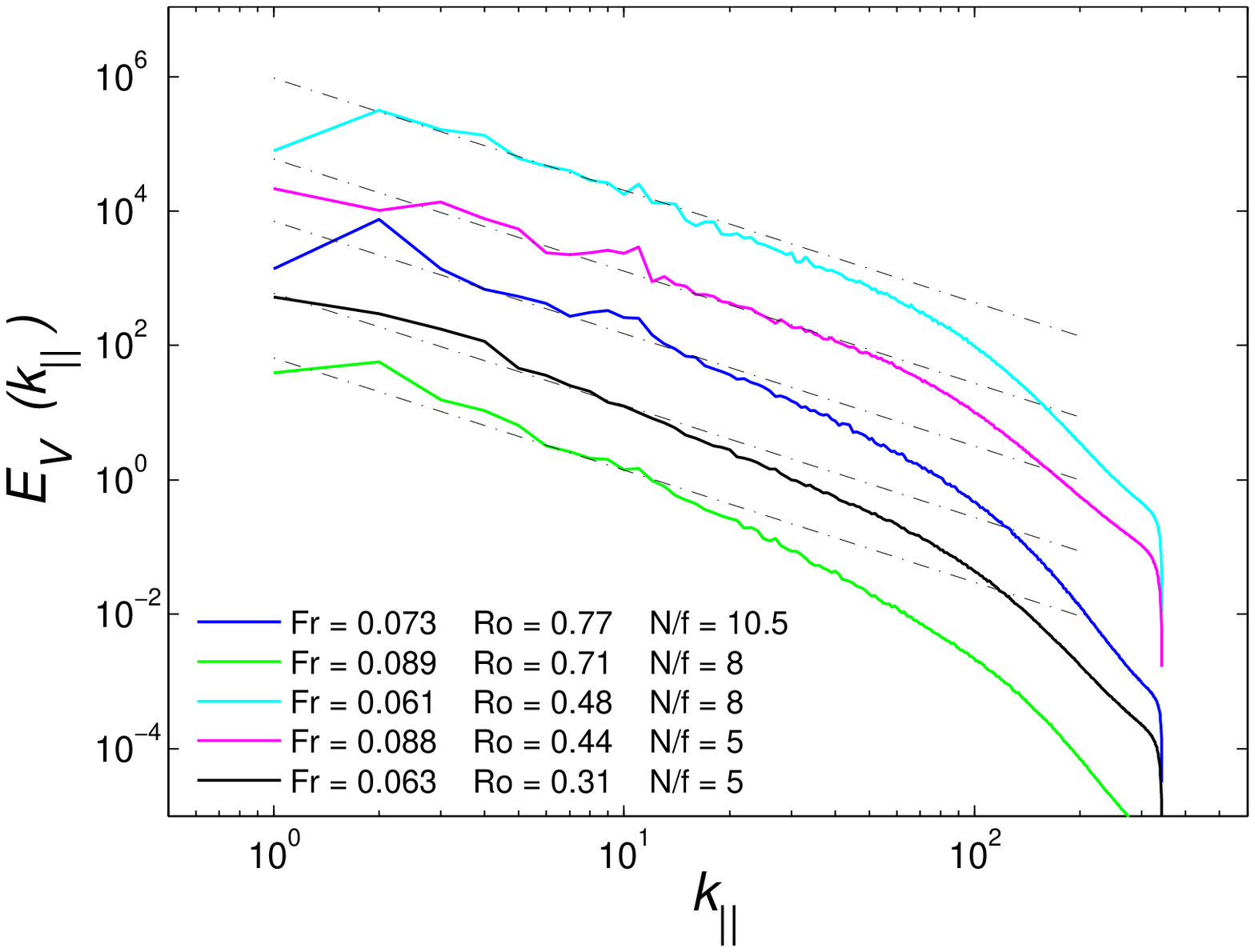}     
\caption{Spectra of the runs at $1024^3$ resolution at the final time (see Fig. \ref{f:april-Ta}).
 The plots are color-coded for different parameters (see inserts): blue, green, cyan, magenta and black correspond respectively to run 5, 7, 3, 6 and 4 of Table \ref{tab_Y}, with spectra  shifted by a factor of 10 each, for clarity.
Kinetic (left) and potential (right)  energy spectra in terms of isotropic wavenumber, $E_V(k), \ E_P(k)$ (top), and of perpendicular wavenumber, $E_V(k_\perp), \ E_P(k_\perp)$ (middle). {\it Bottom:} kinetic  energy spectra, $E_V(k_\parallel)$ as a function of $k_\parallel$. The dashed lines represent Kolmogorov spectral indices.
} \label{f:april-Sa} 
\end{figure}

We now examine this behavior  in more detail through the distribution with scale of the kinetic and potential energy and their ratio, for the  runs of Table \ref{tab_Y} at the resolution of $1024^3$ points.
In Fig. \ref{f:april-Sa} are given the energy spectra towards the end of the computation, the kinetic energy having reached the gravest mode, $k=1$. The dotted lines represent  $k^{-5/3}$ scaling laws. At large scales, a Kolmogorov spectrum is clearly followed for the kinetic energy, in terms of isotropic wavenumber and of both $k_\perp$ and $k_\parallel$, corresponding to the inverse cascade; note an accumulation of energy at the smallest wavenumber particularly visible in terms of $k_\perp$, which corresponds to the dynamics of the inverse cascade. The potential energy spectrum also follows a Kolmogorov law in terms of $k_\perp$, showing a strong energy exchange between kinetic and potential modes, but the isotropic $E_P(k)$ spectra (top right) are markedly different, with some growth at scales slightly larger than the forcing scale, followed by a rapid decrease at even larger scales. {Note that the growth of $E_P(k)$ at large scales is similar for all runs except for the one with the lowest Rossby number (in black), for which it is stronger.}

At small scales, the departure from a $-5/3$ scaling for some runs is particularly visible in the kinetic energy spectrum in terms of $k_\perp$, although all 5 runs have comparable dimensionless parameters, with $0.061 \le Fr \le 0.089, \ 0.31 \le Ro \le 0.77,\ 5 \le N/f \le 10.5, \ 28 \le {\cal R}_B \le 57$. In fact, some of these spectra appear to be steeper than $k^{-3}$, and the steep spectra result from the highest $N/f$ values, and represent for this series of runs the highest  Rossby numbers ($Ro\ge 0.48$): stratification prevails and, with low Froude numbers, spectra are steep. On the other hand, the two remaining runs -- runs 1 and 4-- have lower Rossby number and the spectral dynamics is known to be shallower in that case \cite{mininni_12}.
Thus, a possible differentiation in these runs is feasible in terms of the Rossby number being smaller or larger than a critical value estimated to be close to  $0.45$, corresponding to the predominance or not of the inverse cascade. For example, the shallower small-scale energy spectra in terms of $k_\perp$ obtain for the two runs with the stronger rotation (magenta and black curves, $Ro=0.31$ and $0.44$). 

Fig. \ref{f:april-Sb} gives the ratio of kinetic to potential energy as a function of isotropic (left) and perpendicular (right) wavenumbers, with, at left, an insert giving a blow-up of the main part of the inertial range. These flows have a similar behavior, with a strong dominance of the kinetic energy in the large scales, as expected since the inverse cascade is primarily driven by the so-called slow mode, that is by the horizontal velocity. There is a large range of wavenumbers, roughly for $4 \le k \le 100$, where the kinetic and potential energy are in approximate equipartition. 
This quasi-equipartition, noticeably, goes through the region of forcing, with $k_F\approx 10$, towards larger scales as well. It will be of interest to see if such an equipartition depends on time: does it extend to even larger scales as time evolves?

When examining the blow-up of that spectral region, it is seen that the equipartition solution is in fact  being relaxed progressively, 
{at a Froude-dependent wavenumber, after which}
 the relative kinetic energy slowly increases with $k$, in similar fashion for all runs. Finally, there is a steep growth of the kinetic to potential  energy ratio in the vicinity of the maximum wavenumber of the runs, quite close to the dissipation wavenumber (see below). 

The dynamics is more complex in terms of $k_\perp$, and in this case the forcing scale is quite clear, although  there is no potential energy forcing. A relaxation of the dominance of kinetic energy occurs both at small and large scales, with for high wavenumber, a change of behavior at a parameter-dependent wavenumber that varies between roughly 40 and 100. Note that a similar slow trend away from equipartition was found in the run studied in \cite{rosenberg_16} at a resolution of $4096^3$ points for the isotropic spectra; the run in 
 \cite{rosenberg_16} has  parameters that are close to run 3 of this paper, except for a  substantially higher Reynolds number ($Re\approx 54000$), and for the absence of forcing. In the study performed in \cite{marino_15w}, this change of regime in the ratio of kinetic to potential energy was associated with a scale wavenumber, $L_R=2\pi/K_R$, corresponding to the buoyancy scale, which is also the scale at which the dominance of waves over eddies is reversed. The parametric study in \cite{marino_15w} finds a scaling $K_R\sim Fr^{-1}$, which is in rough agreement with the present data.
 Finally note that, at either large or small scale, the excess of kinetic energy is never more than a factor of 10, contrary to the case at left in Figure \ref{f:april-Sb} in terms of $|k|$.

Where does the strong tendency for dominance of the kinetic modes over the  scalar modes in terms of isotropic wavenumber come from?
One can argue that, in the small scales, the temperature or density perturbations are 
{becoming progressively} passive, granted the buoyancy Reynolds number is larger than unity, which is the case for all runs of Table \ref{tab_Y} examined here. One could resort to an argument linked to the behavior of small-scale turbulence for which, in the idealized setting of homogeneous turbulence using Feynman diagram techniques, is known to develop in most cases a turbulent Prandtl number, defined as the ratio of turbulent viscosity to turbulent diffusivity, slightly smaller than unity: the small-scale velocity modes, on average, dissipate less efficiently than the small-scale passive scalar modes, perhaps because the latter are not constrained by the pressure and develop very fine structures in the form of fronts and filaments \cite{celani_01, sullivan_17, mininni_17}.
Indeed, numerous studies, for example using the Renormalization Group  (RNG) formalism \cite{forster_77, pouquet_78} have evaluated  the turbulent Prandtl number $\nu_{turb}/\kappa_{X,turb}$, where $\nu_{turb}$ is the eddy viscosity created by the turbulent motions of the fluid, and similarly $\kappa_{X,turb}$ is the anomalous diffusivity of a passive scalar $X$, such as a chemical pollutant. 
One can also define parallel and perpendicular effective viscosities in the context of anisotropic flows, with their ratio $\nu_\parallel/\nu_\perp$ of order 2 or more in the case studied in \cite{garnier_81}, applying the RNG for MHD flows in the presence of an imposed uniform magnetic field. 
These studies show that the turbulent Prandtl number is of order unity, but not quite;  phenomenology would suggest that $Pr_{turb}\approx 1$  since these two transport coefficients can both be estimated dimensionally in the simplest case as $\sim U_0L_0$. However, it is found  to be in general slightly smaller than unity, $\approx 0.7$ \cite{forster_77}, with a variation of the order of up to 15\% in the presence of helicity \cite{yousef_03, jurcisinova_14}. Similarly, when considering rotating stratified flows, the turbulent Prandtl number which is an essential parameter when considering anomalous dissipation in the ocean, is again found to be $ \lesssim  1$ for buoyancy Reynolds numbers of the order of 50 or above \cite{salehipour_15}. Such an imbalance in the effective small-scale dissipation could be at the origin of the clear but rather small tendency of the $E_V/E_P$ ratio to increase at small scales, up to quite close to the dissipation wavenumber. 

On the other hand, the steep accumulation of kinetic energy  close to the numerical cut-off may not be explained by such a subtle effect. It is in fact 
systematically found in a separate study of rotating stratified flows in the absence of forcing (see \cite{rosenberg_16} for temporal data), whereas 
the ratio of wave to eddy modal energy, as done in \cite{bartello_95, herbert_16}, does not display this effect \cite{marino_17}. 
This latter fact excludes any numerical origin to this accumulation. Another possibility is the fact that fully helical structures develop at small scale, weakening the nonlinear advection term which is the source of the turbulent viscosity, whereas no such geometrical effect is directly available to the scalar field. This point clearly deserves more study. Furthermore, at these small scales, dissipation sets in and the zero forcing and initial conditions in the potential fluctuations lead to an abrupt and strong transition towards dominance of kinetic energy.

Finally, this steep increase in $r_E(k)$ may  be related to the fact that, at small scales, the fluxes of potential energy for the runs described in this paper are all strongly negative, corresponding to a gain of kinetic energy at small scales (see Fig. \ref{f:april-Fa}, and \S \ref{SS:EF}), 
since the total energy flux is constant. This could be due to small-scale instabilities which at these buoyancy Reynolds numbers are known to exist. Such a complexity in spectral dynamics reflects the complexity of the anisotropic energy exchanges between scales and between energetic modes. 

 \subsection{Energy fluxes} \label{SS:EF}

A clear diagnostic of the dual cascade phenomenon is through the observation of the energy flux. Oceanic data indicates that the energy flux can be  both positive at some small scales, and negative at larger scales \cite{arbic_13}, but the spatial resolution for such data is not very high. Using high-resolution DNS of forced RST with grids of up to $2048^3$ points, it was shown clearly in \cite{pouquet_13b, marino_15p} that the energy flux can be constant and negative at large scales and constant and positive at small scales. This large parametric study allowed for finding a  scaling with dimensionless parameters of these two fluxes, and specifically of their ratio 
\be
R_\Pi=|\Pi_{LS}|/\Pi_{ss} \sim [Ro*Fr]^{-1}
\label{RPI} \ee
where $\Pi_{LS,ss}$ stand respectively for the large-scale and small-scale total energy fluxes.

In the absence of mean flow, for strong waves (or weak eddies), a weak turbulence formalism can be developed leading at lowest order to closed integro-differential equations for the energy spectra \cite{zakharov_92, newell_11, nazarenko_11b}. There are several limitations to this theory. For example, it is non-uniform in scale, since the small parameter -- {\it a priori} the ratio of the wave period to the nonlinear eddy turn-over time, or here the Froude or Rossby numbers -- varies with scale. Moreover, other characteristic times can arise which, at some scale, are faster than the waves, such as the sweeping  by large-scale motions of small--scale eddies \cite{clark_14, mininni_17}. Indeed, it is shown in \cite{winters_94} using numerical modeling that, at a critical layer \cite{sun_15}, roughly one third of the energy is transmitted to the mean flow.
As noted in \cite{newell_11}, ``{\sl For wave turbulence, we are only at the beginning of the experimental stage}.''
These specific issues are discussed more at length in \cite{mininni_17}, in the  context of rotating and/or stratified flows.

\begin{figure}   
                            \includegraphics[width=.45\textwidth]{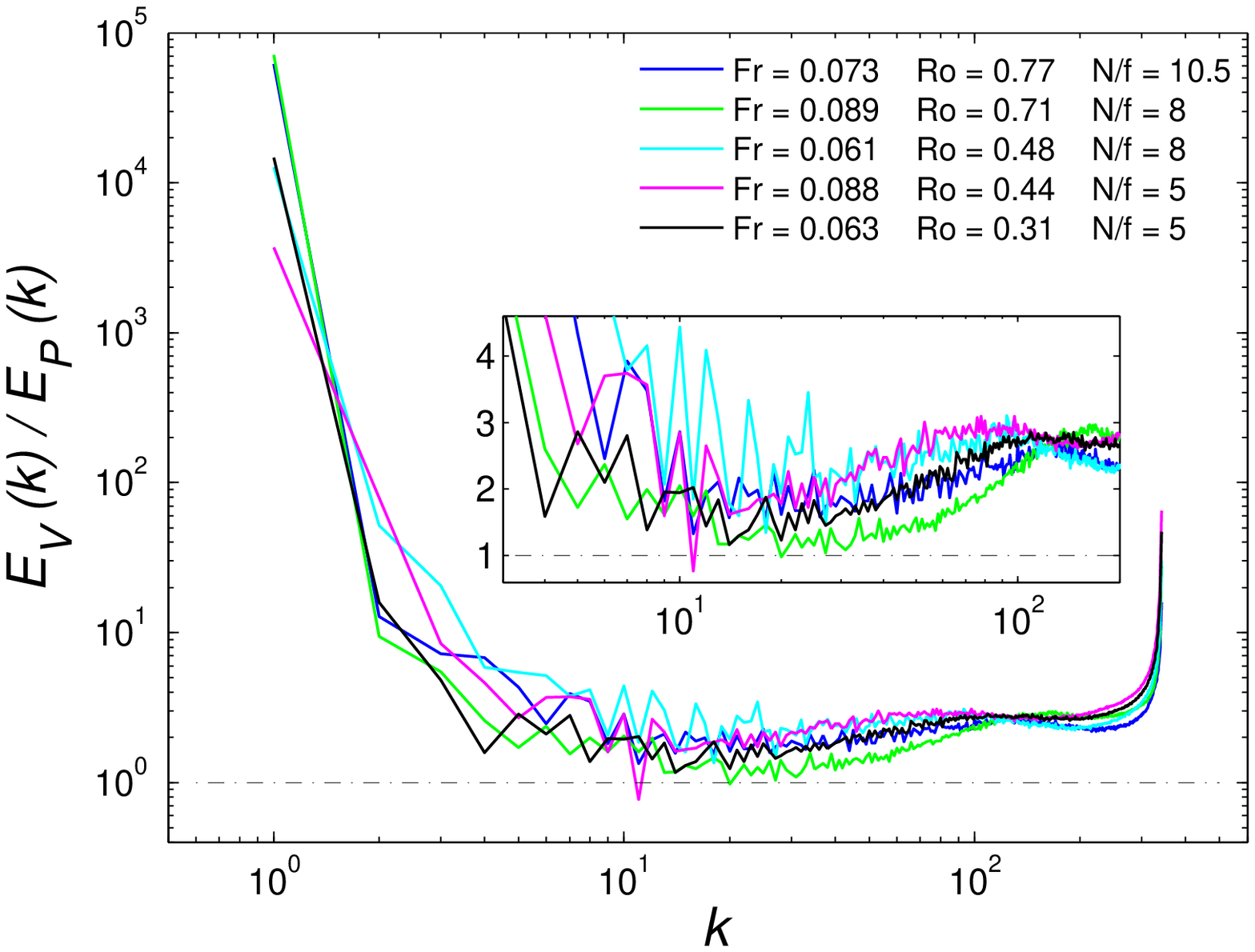}
                            \includegraphics[width=.45\textwidth]{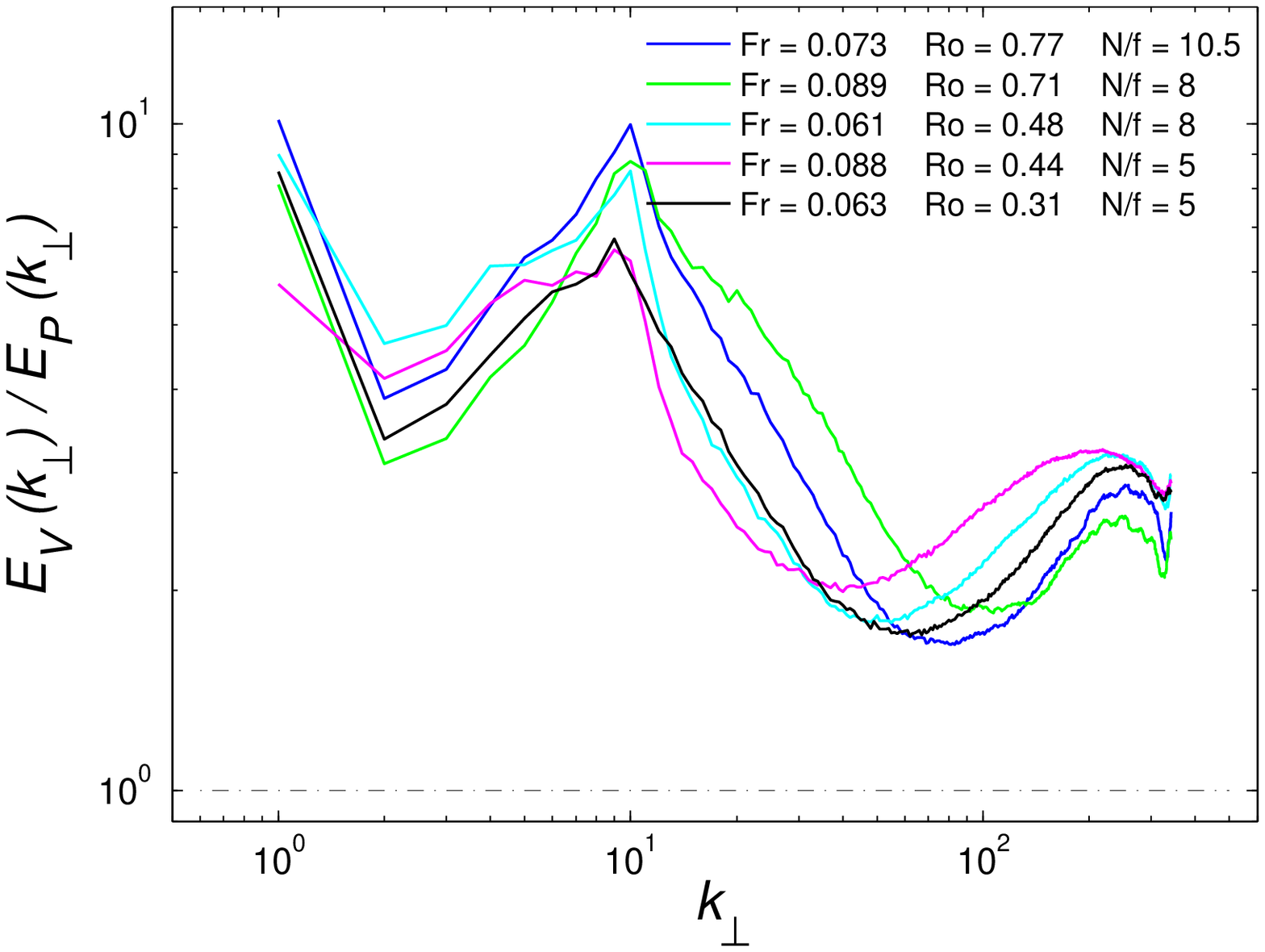}  
\caption{Mode-by-mode ratio of kinetic to potential energy spectra for the runs on $1024^3$ grids  at the final time, in terms of isotropic wavenumber, $r_E(k)=E_V(k)/E_P(k)$ (left), and of perpendicular wavenumber, $r_E(k_\perp)$ (right). The plots are color-coded as in Fig. \ref{f:april-Sa}. The insert at left is a blow-up of the central spectral region. 
{The dash-dot horizontal black lines indicate $E_V/E_P=1$.}
} \label{f:april-Sb} \end{figure}

\begin{figure*}   
 \includegraphics[width=.45\textwidth]{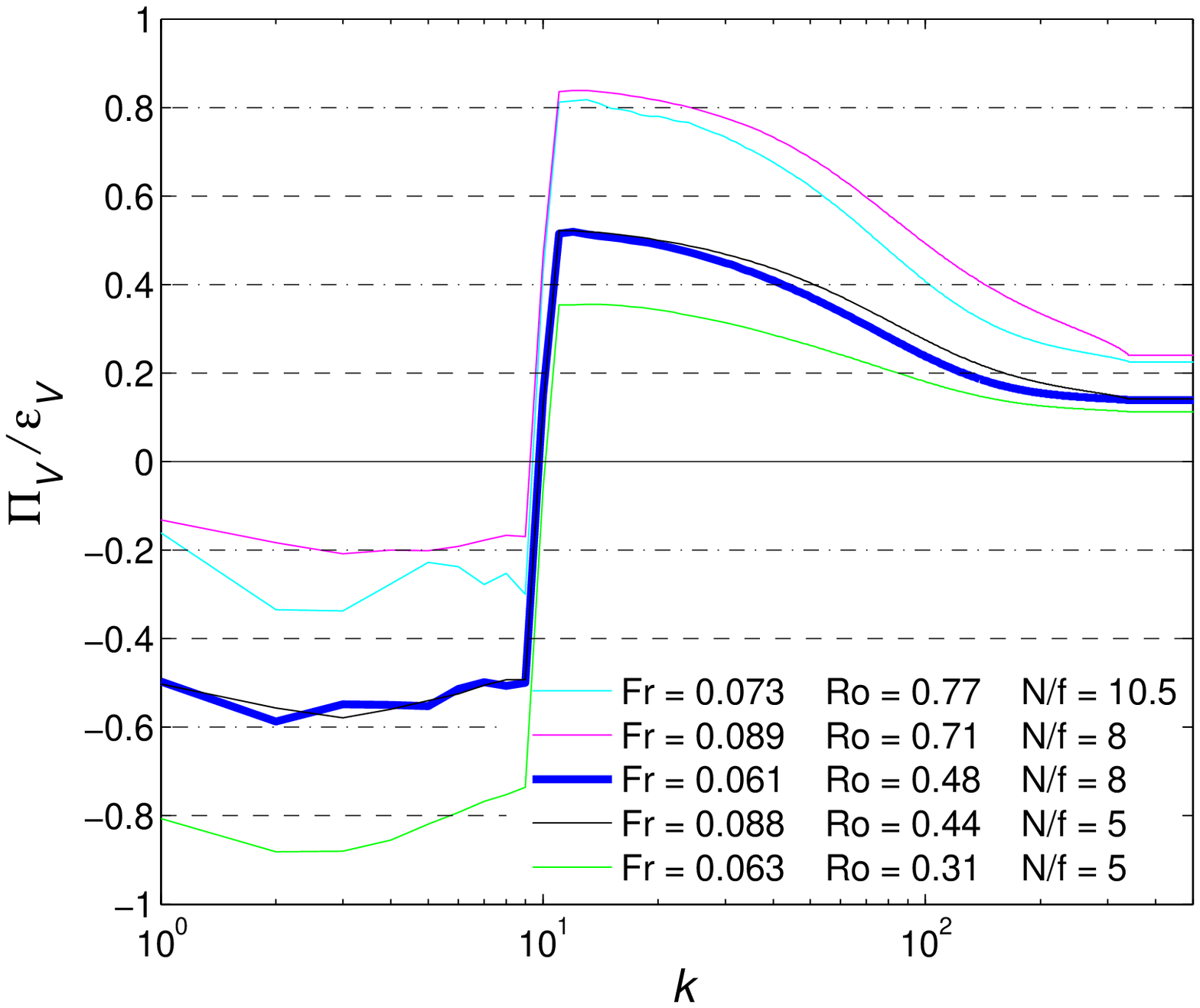}  
                           \includegraphics[width=.45\textwidth]{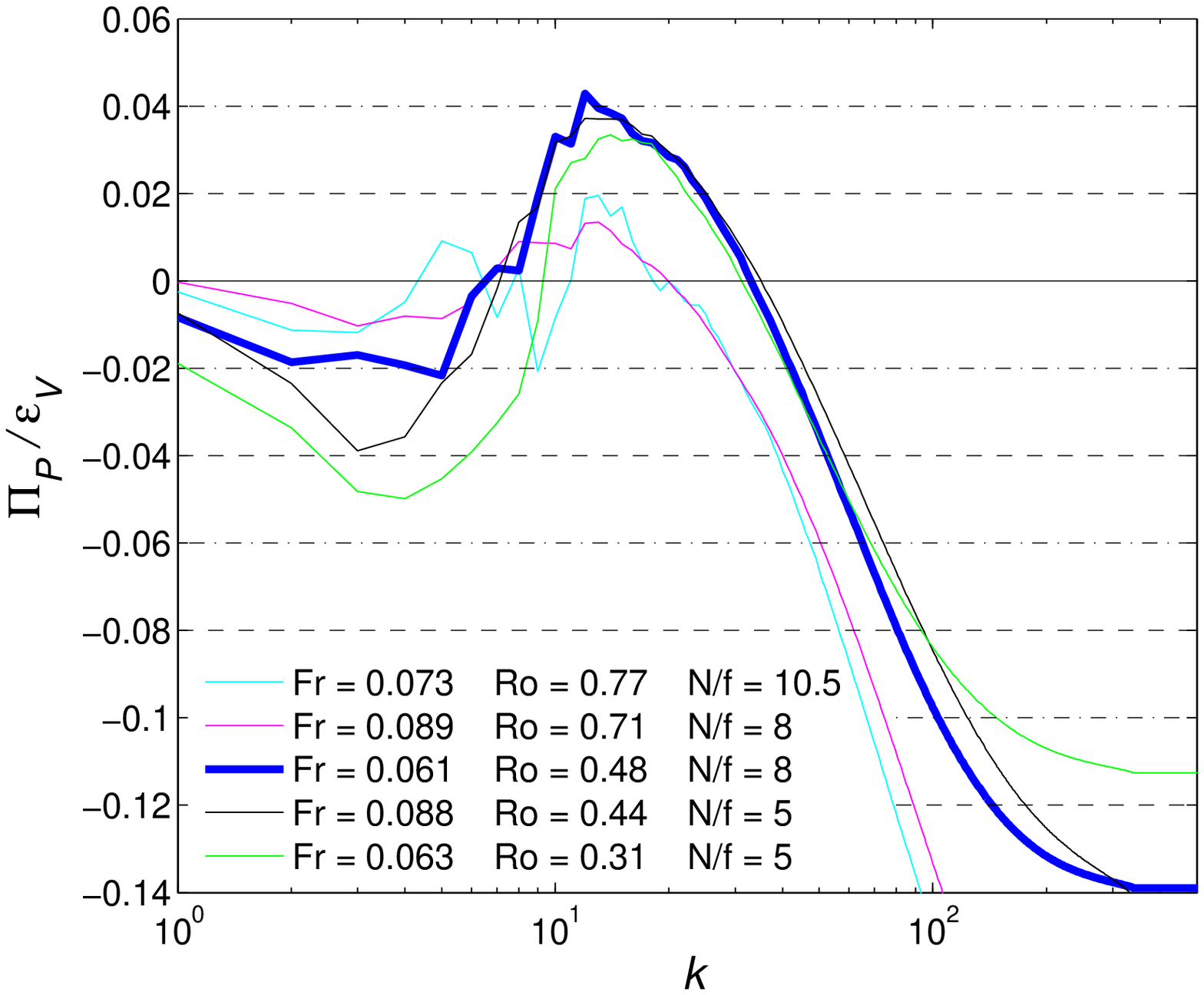}  
                           \includegraphics[width=.45\textwidth]{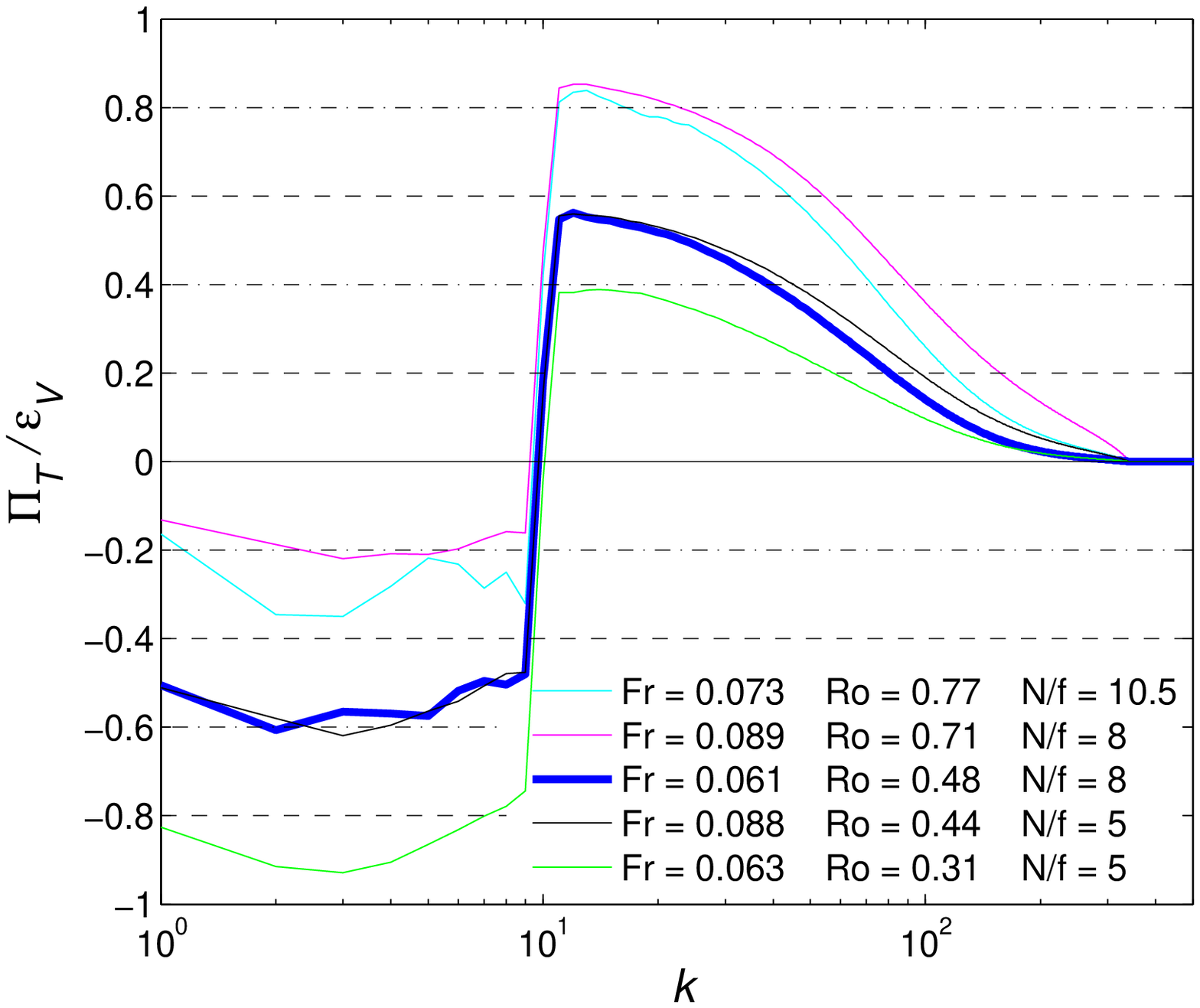} 
 \caption{
Flux spectra for the runs at $1024^3$ resolution in terms of isotropic wavenumber, and all normalized by the kinetic energy dissipation rate $\epsilon_V$, at the final time. 
The plots are color-coded for different parameters (see inserts), with blue, green, cyan, magenta and black now corresponding respectively to run 3, 4, 5, 7 and 6 of Table \ref{tab_Y}. 
} \label{f:april-Fa} \end{figure*}

The result in \cite{marino_15p},  using a phenomenological argument backed by large direct numerical simulations, of a variation with Froude and Rossby number of the ratio of the inverse to direct total  energy flux does not consider directly the possible variations in the forcing amplitude or shape function. Indeed, the forcing ${\bf F}$ is constant in the momentum equation, but results in an energy input that depends on the dynamics of the flow, that is $\left< {\bf u} \cdot {\bf F} \right>$. Therefore, the effective energy input due to the forcing is likely to change across the set of runs. 
For example, in runs dominated by strong stratification, or for the purely stratified case, let us assume that there is a balance between the upscale vortex mode energy (in $k_\perp$), if there is such an upscale cascade, and the so-called saturation energy spectrum  in the small scales, or that $k_\perp E(k_\perp) = k_\perp^{-2/3}\epsilon_V^{2/3}=N^2k_\parallel^{-2}$, where the characteristic vertical scale is the buoyancy scale $U/N$. Expressing $\epsilon_V=\epsilon_D=U^3/L_{int}$, for high-enough buoyancy Reynolds number,  yields $k_\perp\sim k_{int}$, i.e. under these circumstances, no inverse cascade is taking place; in the rotating stratified case, it is known that the ratio of the direct to inverse flux is proportional to $FrRo$ \cite{marino_15p}: in the limit of no rotation ($Ro\rightarrow \infty$) at fixed Froude number, the direct energy flux is dominant, consistent with the previous argument.

 The flux ratio scaling given above is explained on a phenomenological basis relying on the fact that the turbulence in these cases is weak, and that the inverse cascade is stronger when the rotation is stronger. 
This scaling is not well observed when the buoyancy Reynolds number is not sufficiently large for the turbulence to be developed,
and a different scaling emerges for $N/f\approx 8$. This latter result is related to the fact that for these values, the Rossby number is too large for an effective inverse cascade to develop, as noted before,  since the Froude number must be large enough to have ${\cal R}_B\ge  20$, and yet still small enough to be in a regime still dominated by the wave dynamics at large scales. 

We show in Fig. \ref{f:april-Fa} the kinetic, potential and total energy fluxes, $\Pi_{V,P,T}(k)$, normalized by the kinetic energy dissipation rate $\epsilon_V$, computed individually for each case. The fluxes are all averaged over 5 to 6 turn-over times. Note the different scale for $\Pi_P(k)$: the potential flux is at best 10\% of the kinetic flux, the energy of the horizontal velocity being the driver of the inverse cascade. $\Pi_V$ shows a sharp variation around the forcing wavenumber $k_F\approx 10$, whereas the change around $k_F$ is smoother for $\Pi_P$.
For the two runs with higher Rossby number, there is as expected a strong direct flux and a weaker inverse flux for $E_V$. For the two runs with comparable Rossby numbers at an intermediate value, the fluxes are comparable, and direct and inverse fluxes are similar.  Finally, the run with the smaller $Ro$ has the strongest inverse flux. The run with $Ro=0.48$ (thick blue line) is intermediate between high and low Rossby numbers. For $Ro<0.48$, the behavior is different, with now more inverse than direct kinetic energy flux. 

The behavior of the potential energy flux is more complex (see Fig. \ref{f:april-Fa}, top right). At small scales, it is strongly negative, leading to an input of kinetic energy at those scales. This may be linked to the excess of kinetic energy in the very small scales observed in Fig. \ref{f:april-Sb}. It corresponds to the many instabilities that can arise in such flows, as overturning occurs because of vertical shear, or through frontal collapse \cite{mcwilliams_16} 
{(see also \cite{mininni_17c})}. 
At large scales, the potential energy flux is negative but rather small. 
Note that the ordering for the inverse cascade flux is the same as for the flux of $E_V$, and is governed by the Rossby number.  For the two larger $Ro$, the inverse and direct fluxes around the forcing wavenumber $k_F\approx 10$ are comparable and small.
Finally the total flux (Fig. \ref{f:april-Fa}, bottom) is quite close to the kinetic energy flux which is dominant, except towards $k=k_{max}$, indicating that indeed the total energy is conserved and the total energy flux has thus to be zero there. 

 \subsection{Exact laws}

Another possible diagnostic for these dual cascade systems is through the use of so-called exact laws for turbulent flows, laws that stem from conservation principles and that relate directly third-order structure functions and the injection (and dissipation) rate of the invariant \cite{K41b, yaglom_49, fulachier_76, antonia_97, augier_12b}. Such laws, obtained under a set of hypotheses, namely incompressibility, stationarity, high Reynolds number, homogeneity and isotropy,  allow one to deduce the actual values of the dissipation rates of energy, as done for example in the Solar Wind \cite{marino_08,marino_11,marino_12}, and potentially in the atmosphere and the ocean as well. One can  also observe directly the sign of the third-order structure function, which is a straightforward indication of the nature of the transfer, upscale or downscale. In the simplest case, that of HIT, the Kolmogorov (1941) law  is written as \cite{K41b}:
$$
\left< \delta u_L^3(r) \right> = -\frac{4}{5} \epsilon_V r \  \ , \ \ \delta u_L(r) = u(x+r)-u(x) \ ;
$$
 the  velocity difference is taken along the distance ${\bf r}$,  $u_L$ being the longitudinal component of the velocity projected on ${\bf r}$. A remarkable feature is that we have both the sign and the numerical factor in this law, but we do not know {\it a priori} the sign of the kinetic energy flux rate $\epsilon_V$.  The corresponding exact Yaglom law for a passive scalar is  \cite{yaglom_49}:
$$
\left< \delta u_L(r) \delta \rho(r)^2  \right> = -\frac{4}{3} \epsilon_P r \ .
$$
It has been tested on observational data in the lower stratosphere at roughly 10 km of altitude, using commercial aircrafts \cite{lindborg_00},  with the assumption that at small scale the scalar is passive, which can be expected {\it a priori} beyond the Ozmidov scale at which isotropy and HIT recovers ($\approx 10\ m$). 
Note that this assumption is not necessarily needed since the extra term in the scalar equation, that of energy exchanges with the vertical velocity, occurs on a time of the order of the inverse \BV, possibly much faster than the sampling frequency. 
A negative flux is found at large scales, with a change of sign around $10^3\ km$; from this, it is deduced that the large scales undergo an inverse cascade and the small scales a direct cascade. Similarly, changes of signs in energy fluxes are found in \cite{jiang_10} observing the troposphere over Owens Valley (California), with the change of sign associated with Kelvin-Helmoltz instabilities leading to wave breaking and thus to small-scale dissipation. In Antarctica as well, the data indicates changes of sign of the flux function, but in this latter case, these changes do not take place simultaneously \cite{king_15a}, and are rather linked to the onset (or not) of convective motions.

In \cite{deusebio_14b}, the Kolmogorov (1941) and the Yaglom (1949) laws are computed using direct numerical simulations for a set of rotating stratified flows at fixed small Froude number, $Fr=0.01$. Two input formulations are used, one forcing only the geostrophic component of the flow, and the other one forcing also the ageostrophic eddies. These laws, in terms of isotropic distance $r=|{\bf r}|$, show a change of sign (indicative a priori of a change of direction of the energy cascades) at different scales between themselves and for different $N/f$. Such a  sign change is associated with the geostrophic component of the flow.
Flux-laws with at least one sign change in the third-order structure function  are also observed in fact in numerical models of the atmosphere \cite{waite_09}.
These same exact laws will be computed for the runs described in this paper in the near future.

\section{Geophysical turbulence and anisotropy} \label{S:aniso}          

The recovery of small-scale symmetries in HIT, such as isotropy,  has been studied  for a long time, for example finding a quasi-linear relaxation to isotropy for axisymmetric flows, using  either the Direct Interaction Approximation (DIA)  \cite{kraichnan_59} or DNS \cite{herring_74}. As mentioned earlier, since it is known that in HIT the helicity has a $k^{-5/3}$ spectrum (omitting possible intermittency corrections), the relative helicity, $H_V(k)/[kE_V(k)]$ decays slowly and full isotropy with negligible helicity is only recovered as $1/k$.

As reviewed in the introduction, the quasi bi-dimensionalization of flows implies strongly anisotropic statistics for such configurations, a topic tackled in numerous instances, and more recently with improved computer power.
For purely stratified flows, it was found,  in the context of turbulence collapse, that vortices and waves co-exist on different time-scales \cite{hopfinger_87}. Anisotropy in rotating turbulence was studied in \cite{cambon_89} using the Eddy-Damped Quasi-Normal Markovian (EDQNM) closure model including the Coriolis frequency $f=2\Omega$ in the eddy-damping of triple correlations; these authors could show a trend towards bi-dimensionality insofar as there was a transfer from the parallel to the perpendicular wavenumber energy (see also \cite{cambon_04b, sagaut_08} for reviews of more recent works, and \cite{delache_11} for a model of rotating flows). Anisotropic spectra were derived in the context of weak turbulence in \cite{galtier_03}, but with no inverse cascade because of the decoupling, at lowest order, of the 2D modes from the 3D waves, as well as in \cite{cambon_04b} on the basis of a phenomenological argument following weak turbulence scaling, with a prefactor proportional to $\Omega^{1/2}$ (see also \cite{cambon_94}).  Anisotropic structures correspond to Taylor columns for purely rotating flows,  to horizontal  layers in purely stratified flows, and such layers become progressively slanted as rotation is increased, with {\it a priori} an isotropic flow for $N=f$ since the dispersion relation becomes simply in that case, $k \sigma(k)=N=f$.

In RST, the large-scale anisotropy and the small-scale isotropy, provided the buoyancy Reynolds number is large enough,
 can be diagnosed respectively through the velocity and vorticity statistics, as well as the temperature field and its gradients. This is done for example in the context of the Rayleigh-Taylor instability, with small-scale properties, measured in the middle of the mixing layer, being similar to HIT \cite{cabot_06, cabot_13}.
Another simple way to diagnose anisotropy is through the ratio of the perpendicular to parallel integral scales and Taylor scales (see eq. (\ref{length3}) for definitions), $r_L=L_\perp/L_\parallel, \ r_\lambda= \lambda_\perp/\lambda_\parallel$, the former  characterizing large scales and the latter the small scales.
For rotating flows, $r_L$ is found to vary by a factor three between weak and strong rotating cases \cite{mininni_09c, mininni_10a}, whereas $r_\lambda$ itself, sensitive to small scales, is quite close to unity.  

Furthermore, in regimes in which the nonlinear part of potential vorticity, namely $\vomega \cdot \nabla \theta$, can be neglected, it is predicted in \cite{kurien_08} that $E_P\sim k_\parallel^{-3}$ for $k_\parallel >> k_\perp$ i.e. for pancakes as the predominant structure, whereas in the opposite case of a columnar regime, it is the energy spectrum of the perpendicular velocity field that becomes proportional to $k_\perp^{-3}$. This is a finding that can be seen as a generalization of that of Charney \cite{charney_71} when moving away from geostrophic balance. This prediction, backed up by computations done for $N/f=1$, is based on the dominance of the potential enstrophy cascade at small scales (also observed for some cases in \cite{marino_15p},  Supplementary material). This enstrophy cascade is interpreted as being an anisotropic constraint of the conservation of potential enstrophy on energy spectra, but note that for such a frequency ratio, there are no exact resonant interactions \cite{smith_02}. 

It was discussed in detail in \cite{marino_13i, marino_14} (see also \cite{sagaut_08} for thorough reviews) how the transfer of energy occurs in rotating stratified flows as a function of the angle between the wavevector ${\bf k}$ and the  direction of rotation and gravity, assumed to be common and taken as the vertical. The purely stratified  and the purely rotating cases are quite different in this regard, because of the strong anisotropy of the dispersion relation, respectively $k\sigma(k)=Nk_\perp$ and $k\sigma(k)=fk_\parallel$.
In the purely rotating case, in the limit of strong rotation, no transfer occurs in the vertical direction; energy at the forcing scale (or initial conditions) moves towards $k_\parallel=0$; the flow becomes two-dimensional and then undergoes a nonlinear inverse cascade of energy, as originally envisaged by Kraichnan, in the so-called 2D vortex  mode, with no vertical velocity. The transfer in terms of $k_\perp$ is constant and negative; however, the flux in terms of $k_\parallel$ is negligible. 

 Finally, note that anisotropic flows can be modeled through a generalization of two-point closure equations introduced in the isotropic case (see \cite{sagaut_08} for review), using analytical formulations  of both eddy viscosity and eddy-noise, in terms of energy and helicity spectra that are dependent on scale and on time. This is done for rotating flows in \cite{baerenzung_10}, in a framework that is compatible with the model developed in \cite{chollet_81}. A recent work incorporating the effect of small-scale helicity in both the turbulent viscosity and turbulent noise can be found in \cite{baerenzung_11}; this modeling allows for the confirmation of the indetermination (untanglement) of spectral indices for the energy and helicity, as discussed earlier. Models of helical shear flows using helicity as an essential ingredient were also developed in \cite{yokoi_93}. Helicity can lead to the generation of large-scale flows, similar to the MHD case of the helical dynamo, but with the added feature of anisotropy due to solid-body rotation \cite{yokoi_16}, although it is not clear whether such a large-scale instability leads to an inverse cascade proper, a point which will be investigated in future work.

\section{Discussion and Conclusions: The Rich Dynamics of Complex Flows} \label{S:discu}
\subsection{{Summary of new results}}
{
The runs analyzed in this paper have in common the fact that they display a measurable flux of potential energy to the large scales, together with a flux to the small scales, similarly to the kinetic energy, and of course the total energy. Differences between these flows, such as the rate of growth of kinetic and potential energy and dissipation, or the amplitude of fluxes, can be directly attributed to the Rossby number which governs the existence of the inverse energy cascades. 
We also note that, even though the buoyancy Reynolds numbers are not very high in these runs (${\cal R}_B\le 58$), the energy spectra seem compatible with a $-5/3$ index at large scale, consistent with all known studies of inverse energy cascades in the absence of long-time finite-size effects or of large-scale friction. The dynamics of small scales differs from flow to flow, and with angular variations as well. In some cases a -5/3 spectrum is obtained, whereas in other cases steeper spectra appear which could be related to the so-called saturation spectrum resulting from a balance between advection and buoyancy flux in the purely stratified case, but which can also occur here since the Coriolis force does not affect directly  the energy balance. 
\\
There is in fact little variation in Froude number for the runs, of the order of 15\%, thus the changes in spectral behavior at small scale can be related to the role rotation plays in small-scale dynamics in these flows, as well  as to the amount of anisotropy induced both by the rotation, the stratification and their relative strength. This is particularly true considering that the Froude numbers of all the runs considered here are in the so-called intermediate range where waves and non-linear eddies are competing with each other \cite{pouquet_17}. Strong rotation is clearly seen affecting the overall ratio of potential to kinetic energy and potential to kinetic dissipation.
Finally, the scale-by-scale ratio of kinetic to potential energy, shown in spectral space, is found  to settle for all flows, throughout most of the inertial range, to a constant value close to one, 
and with differences in the large scales corresponding to global parameters. 
}

\subsection{{How much dissipation and mixing?}}

The lack of universality in wave turbulence, corresponding for example to different  spectral indices for, say, the total energy, when all the invariants of the problem are the same, is the signature of a richer set of dynamics than that first envisaged in the weak turbulence formalism. This is well documented, for example  for gravity-capillary waves \cite{herbert_E_10}, or for elastic waves \cite{yokoyama_13, miquel_13}, as well as in MHD \cite{lee_10}. Different branches in the ($\sigma(k),k$) diagram are identified for different forcing intensities, and are attributed to the formation of coherent structures  as suggested for example in \cite{newell_08}. Similarly, for rotating or stratified flows, the existence of different branches in that diagram correspond to the interactions of the quasi-linear waves with a mean flow, which  may have been produced by the nonlinear interactions in the fluid  \cite{clark_14, clark_15, mininni_17}. A similar phenomenon is already known in turbulent flows that are inhomogeneous and anisotropic. For example,  the shear layers of a von K\'arm\'an flow  are shown to be destabilized, as with a Kelvin-Helmoltz instability, leading to enhanced dissipation at small scale \cite{cortet_09, herbert_E_14}, a phenomenon directly observed in Antarctica in the Romanche fracture, with the strong turbulent mixing  directly linked to long KH billows \cite{vanharen_14}.
As such, and as already envisaged in \cite{hussain_70} when examining a wave in a turbulent shear flow, the velocity field should be decomposed into a mean flow $U$ and a fluctuating component $u^\prime$, the latter having two components itself, a coherent or wave part $u_c$ or $u_w$, and a purely stochastic turbulent part, $u^{\prime\prime}$. Thus, a spatio-temporal analysis should be performed for these flows, with the turbulent flow viewed as a perturbation to the time-averaged flow, in which the pure waves should disappear. As suggested in \cite{herbert_E_14}, the sequence of bifurcations at low and at high Reynolds number appear to be similar, leading these authors to the conclusion that this provides a strong justification to the approach of modeling turbulent flows with eddy viscosities.

 When examining the set of decaying runs analyzed in \cite{rosenberg_16, pouquet_17}, the kinetic energy decay rate $\epsilon_V$ varies by one order of magnitude when the Froude number varies in the range $10^{-3}\le Fr \le 0.26$; a large range of  $N/f$ values is covered in this study; on the other hand, when keeping $N/f$ constant, viz. $N/f=4.95$, the energy dissipation rate varies in the range $0.19 \le \epsilon_V \le 0.57$ for $0.04 \le Fr \le 0.14$. For both sets of runs, the Reynolds number is rather constant (between $8\times 10^3$ and $1.4\times 10^4$). In both cases, with or without forcing, the variations in $\epsilon_V$ may be the signature of a strengthening of nonlinear interactions compared to the waves, as the characteristic dimensionless parameters increase. The rate of dissipation in a given flow may also depend on the nature of the forcing. For example, it is shown in \cite{mininni_17c} that, in a container with a large aspect ratio, as in the atmosphere but with a more modest value, namely $L_{0z}/L_{0\perp}=8$ where $L_{0z, 0\perp}$ are characteristic length scales in the vertical and the horizontal, leading to the presence of strong large-scale shear layers, the energy dissipation can be close to the dimensional evaluation, namely $\epsilon_D=U_0^3/L_{0\perp}$, even for buoyancy Reynolds number that are moderate (specifically, ${\cal R}_B\approx 30$ in the computation in \cite{mininni_17c}); this energetic flow dissipation is associated with the formation of fronts and filaments, as already examined in \cite{mcwilliams_16} (see also \cite{sullivan_17}).
{Furthermore, several observations and numerical simulations indicate that, for  stratified turbulence in the presence of weak rotation, the mixing efficiency $\Gamma_f$, that is the ratio of the buoyancy flux $\left< N w \rho \right>$ to the kinetic energy dissipation rate, decreases as $Fr^{-1}$, as detailed in \cite{pouquet_17}; it  can be related to a ${\cal R}_B^{-1/2}$ scaling at fixed Reynolds number, assuming $Re$ is high enough, a scaling rather commonly observed (see {\it e.g.}, \cite{mater_14}). This is because 
in the  regime of  stably stratified strong turbulence,  the density fluctuations have become passive, and isotropy has recovered so that $w\sim U_0$ and $\beta\approx 1$, leading to the desired scaling.
 }

However, many phenomena combine to render the weak turbulence theory invalid.
One concerns finite box effects, as demonstrated for example in \cite{kartashova_08}: four-wave resonances can form independent clusters of Fourier modes, hence weakening the energy transfer to (small) scales, and thus leading to steeper energy spectra.
As argued in \cite{nazarenko_09},  taking the limit of an infinite box must precede taking  the limit of the small parameter leading to a closure of the equations: in other words, the resonance broadening must be larger than the discretization step in Fourier space due to the limited length of the computational box or experiment. Another argument  concerns the non-uniformity in scale of the assumption of a small parameter for the weak turbulence closure to be valid.
{It was shown recently, using  laboratory experiments  \cite{campagne_15} as well as direct numerical simulations \cite{clark_15}, that the sweeping of turbulent eddies by large-scale flows can also alter significantly the effective dispersion relation as documented in a spatio-temporal, $\omega_k-k$ diagram; thus, it modifies the break-down from the weak  wave turbulence to the strong turbulence regime, thereby possibly altering the lateral and vertical mixing and dissipative properties of such flows which are central to an accurate modeling of geophysical flows for weather and climate.
}

\subsection{{Two-dimensional versus three-dimensional behavior,  intermittency and numerical fractality}}

It was shown in \cite{frisch_76} that the transition between a 2D and a 3D behavior, in a cubic  box for non-helical flows, could be modeled through an analytic continuation of the equations (in Fourier space) to a variable, real dimension, $d\in {\cal R}$. A cross-over between the 2D and 3D behavior is found using a numerical integration of a model for a critical dimension of $d_c\approx 2.03$, whereas in \cite{kraichnan_76b}, the critical dimension is evaluated on the basis of a change of sign of the eddy viscosity computed through the use of the Test Field Model \cite{kraichnan_71}. This was more recently suggested heuristically in \cite{smith_96} for rotating turbulence (see also \cite{celani_10}). Similarly, it was found in \cite{frisch_12} using a fractal Fourier decimation for the inverse cascade, that the inverse cascade existed down to a critical dimension $D_c=4/3$ (see also \cite{lvov_02}), with a diverging Kolmogorov constant.
On the other hand, using again a  fractal Fourier decimation, it was shown in \cite{lanotte_15, lanotte_16} that the flow loses abruptly its intermittency -- observed {\it via} a decrease of the extent of the wings of probability distribution functions of velocity gradients -- and multi-fractality -- observed {\it via} a decrease of skewness and flatness, with the kurtosis of the vorticity going from $\approx 11$ for $d=3$ to $\approx 4$ for $d=2.98$ -- 
as soon as the effective dimension of the set of Fourier modes goes below $d_{cc}\approx 2.98$. 

In that light, it has been stated that the large scales in rotating stratified turbulence behave like a 2D flow, in quasi-geostrophic equilibrium, whereas the three-dimensional small scales are under the influence of gravity waves, such flows realizing naturally a transition between 2D and 3D behavior, with a change at the Zeman scale \cite{mininni_12, biferale_16} or at the Ozmidov scale (see \cite{sagaut_08} for review). This result may be of use in the following context: for a well-resolved direct numerical simulation, one expects that the Kolmogorov dissipation scale be twice the minimally resolved scale. This means that in fact most of the modes are in the dissipative regime, which could be modeled in simpler ways such as a progressive decimation \cite{meneguzzi_96}, or using a stochastic elimination of modes, or a form of eddy noise \cite{palmer_12}. For example, in \cite{fathali_17}, it is shown that the ratio of turbulent production to turbulent dissipation remains approximately the same, down to an effective dimension of $\approx 2.7$, corresponding to the elimination of more than one quarter of the total number of modes, although each term varies, and in particular the angular distribution of energy is modified. This tendency towards a bi-dimensionalization of the flow is associated, as expected, with the growth of an inverse cascade together with a diminution of intermittency. It would be of interest to re-tackle these studies of random decimation of modes for flows with large aspect ratio, to see whether such results remain in that case. 

Large-scale anisotropy has been attributed to truncation errors \cite{fang_17} at low wavenumbers, and is sensitive to the form of the energy input, or rather to the ratio of the scale at which energy is fed into the system to the overall size of the container. Given the computational cost to increase such a ratio, it is further suggested in \cite{fang_17} that a damping be introduced at large scales, a point that will require further study. An inadequate simulation of the isotropy of the large scales can also  lead to spurious energy decay, as clearly shown in two dimensions \cite{chasnov_97}.
On the other hand, at small scale, the residual anisotropy, as such might be the case in HIT simulations at Reynolds numbers that are not quite high enough, has been shown recently to affect differently the scaling of longitudinal and of transverse structure functions, an effect which disappears at higher Reynolds number, as shown on data using a grid of $8192^3$ points, with a Taylor Reynolds number of $\approx 1300$. 
 
 \subsection{{Concluding remarks}}

Turbulence is more complex than was thought 50 years ago, with richer dynamics brought about by the variety of time scales and length scales that can characterize and affect the dynamical evolution of such flows. From entrainment, {\it i.e.} the sweeping of structures by the large-scale field, to untanglement, {\it i.e.} the effect of advection by the velocity field of all field variables leading to non-universality of spectra, new dynamics arise that lead to a richness of behavior of such flows, that should still be considered as turbulent, 
{but in the presence of waves,}
 with multi-scale interactions, constant flux scaling behavior, multi-fractality and intermittency, and with physical insight gained from statistical mechanics centered on the role of the ideal invariants preserved by the truncation.

The path, from the seminal work of R.H. Kraichnan on the inverse cascade of energy in two-dimensional fluids, has proven to be quite rich and varied, with extensions to many physical environments.  Much work remains to be done in the study of flows with dual cascades and with anisotropy, as for instance the 
eventual role of the type of  initial conditions and forcing (balanced or not, constant or variable in time, large scale or small scale), and what are the different regimes, in terms of buoyancy Reynolds number, Reynolds number or  Rossby number, as well as the resolution by the flow of some of the characteristic length scales ($L_B,\ L_{Ell}, \ \ell_{Oz},\ \ell_{Ze}, L_D$). 
For example, it is known that mixing efficiency depends on ${\cal R}_B$  \cite{ivey_08, mater_14b, karimpour_15}, with a weak dependence on rotation {in the decay case} \cite{rosenberg_16, pouquet_17}, but such mixing should also be studied in the dual cascade framework as described here. This is left for future work.
Improved understanding of such flows, as for example in the interchange of energy between kinetic and potential modes \cite{galperin_10, zilitinkevich_13}  will lead to better parametrizations of unresolved small scales in large weather, oceanic and climate codes, allowing for a better modeling of the atmosphere and the oceans. 

\vskip0.5truein
\begin{acknowledgments}
Bob Kraichnan was always a strong supporter of the turbulence research performed at NCAR in the context of atmospheric and oceanic flows, and was a regular visitor there. His help to the turbulence team at the time was invaluable, and he is sorely missed. \\
This work  analyzes runs which used resources of an ASD allocation at NCAR which is  gratefully acknowledged.
Support for AP, from LASP and in particular from Bob Ergun, is gratefully acknowledged as well. RM acknowledges financial support from the program PALSE 
({\it Programme Avenir Lyon Saint-Etienne}) of the University of Lyon, in the frame of the program {\it Investissements d'Avenir} (ANR-11-IDEX-0007). 
PDM acknowledges financial support from UBA-CYT Grant No. 20020130100738BA, PICT Grants Nos. 2011-1529 
and 2015-3530 and the CISL visitor program at NCAR.
\end{acknowledgments}  

\bibliography{ap_17_jun19}    
\end{document}